\documentclass[12pt]{iopart}
\usepackage{graphicx}
\usepackage{amssymb}
\usepackage{xspace}
\usepackage{cite} 

\usepackage{bbm}
\newcommand{\mathds}[1]{\mathbbm{#1}}

\usepackage{color}

\newcommand{\ket}[1]{| #1 \rangle}
\newcommand{\bra}[1]{\langle #1 |}
\newcommand{\rb}[1]{\left( #1 \right)}

\newcommand{\ew}[1]{\left\langle #1 \right\rangle}
\newcommand{\beq}{\begin{eqnarray}}
\newcommand{\eeq}{\end{eqnarray}}

\newcommand{\svec}{\mbox{\boldmath$\sigma$}}
\newcommand{\op}[2]{| #1 \rangle \langle #2 |}

\newcommand{\eq}[1]{Eq.~(\ref{#1})}
\newcommand{\fig}[1]{Fig.~\ref{#1}}
\newcommand{\secref}[1]{Sec.~\ref{#1}}
\renewcommand{\etal}{{\em et al.}\xspace}
\newcommand{\inm}{ideal negative measurement\xspace}

\newcommand{\citer}[1]{{Ref.~\cite{#1}}}
\newcommand{\A}[1]{{\bf (A#1)}\xspace}

\begin{document}

\review[Leggett-Garg Inequalities]{Leggett-Garg Inequalities}

Draft: \today

\author{Clive Emary}
\address{
  Department of Physics and Mathematics,
  University of Hull,
  Kingston-upon-Hull,
  HU6 7RX,
  United Kingdom
  }

\author{Neill Lambert}
\address{CEMS, RIKEN, Saitama, 351-0198, Japan}

\author{Franco Nori}
\address{CEMS, RIKEN, Saitama, 351-0198, Japan}
\address{Physics Department, University of
Michigan, Ann Arbor, MI 48109-1040, USA}

\begin{abstract}
  In contrast to the spatial Bell's inequalities which probe entanglement between spatially-separated systems, the Leggett-Garg inequalities test the correlations of a single system measured at different times.
  Violation of a genuine Leggett-Garg test implies either the absence of a realistic description of the system or the impossibility of measuring the system without disturbing it.
  Quantum mechanics violates the inequalities on both accounts and the original motivation for these inequalities was as a test for quantum coherence in macroscopic systems.
  The last few years has seen a number of experimental tests and violations of these inequalities in a variety of microscopic systems such as superconducting qubits, nuclear spins, and photons.
  In this article, we provide an introduction to the Leggett-Garg inequalities and review these latest experimental developments.  We discuss important topics such as the significance of the non-invasive measurability assumption, the clumsiness loophole, and the role of weak measurements.  Also covered are some recent theoretical proposals for the application of Leggett-Garg inequalities in quantum transport, quantum biology and nano-mechanical systems.
\end{abstract}

\maketitle

\section{Introduction\label{SEC:intro}}

Extrapolating the laws of quantum mechanics up to the scale of everyday objects, one inevitably arrives at the prospect of {\em macroscopic coherence}, with objects composed of very many atoms existing in quantum superpositions of macroscopically very different states.  Schr\"odinger's cat \cite{Schroedinger1935}, simultaneously both dead and alive, is the embodiment of macroscopic coherence.
Needless to say, such a situation runs totally counter to our intuitive understanding of how the everyday, macroscopic world works.

In their 1985 paper \cite{Leggett1985}, Leggett and Garg were interested in whether
macroscopic coherence could be realised in the laboratory and, if so, how one might go about demonstrating its presence.  They approached this by first codifying our intuition about the macroscopic world into two principles:
\A{1} {\em macroscopic realism} (MR) and \A{2} {\em Noninvasive measurability} (NIM).  MR implies that the performance of a measurement on a macroscopic system reveals a well-defined pre-existing value (``Is the flux there when nobody looks?'' \cite{Leggett1985} is thus answered in the affirmative);  NIM states that, in principle, we can measure this value without disturbing the system.
Whilst classical mechanics conforms with both of these assumptions, quantum mechanics certainly does not --- the existence of a macroscopic superposition would violate the first, and its quantum-mechanical collapse under measurement, the second.

Based on these assumptions, Leggett and Garg went on to derive a class of inequalities \cite{Leggett1985} that any system behaving in accord with our macroscopic intuition should obey.  These are the Leggett-Garg inequalities (LGI) and they are the subject of this review.
Should it be shown that a series of measurements on a system violates a LGI, then one of the above assumptions must be invalid and an intuitive macroscopic understanding of the system must be abandoned.
In this way, the LGIs provide a method to investigate the existence of macroscopic coherence and to test the applicability of quantum mechanics as we scale from the micro- to the macro-scopic world \cite{Leggett2002}.

The simplest LGI is constructed as follows.  We assume that it is possible to define for the system a macroscopic dichotomic variable $Q=\pm 1$ and measure its two-time correlation functions $C_{ij} = \ew{Q(t_i)Q(t_j)}$.
We then perform three sets of experimental runs to measure three different $C_{ij}$ with different pairs of time arguments.  Postulates \A{1} and \A{2} together imply that there exists a single joint probability distribution that is sufficient to describe all three experimental runs.  From this it follows that
\beq
    K_3 \equiv C_{21} + C_{32} - C_{31} \le 1
  \label{K3intro}
  .
\eeq
By considering a quantum model of a two-level system undergoing coherent oscillations between the states with $Q=\pm1$, it is easy to show that quantum mechanics violates this inequality with a maximum value of $K_3^\mathrm{max}=3/2$ for the two-level system.

LGIs share the same structure with, and are intimately related to, Bell's inequalities \cite{Bell2004} (compare \eq{K3intro} with the original inequality of \citer{Bell1964}, see also \citer{Wigner1970}). But, whereas Bell's inequalities place bounds on correlations between measurements on spatially-separated systems, in the LGIs, the separation between measurements is in time. LGIs are for this reason often referred to as temporal Bell's inequalities \cite{Paz1993}.  Both sets of inequalities are founded on realism, but to obtain testable inequalities that are violable by quantum mechanics, realism is cojoined with locality in the Bell's inequalities, and with NIM in the LGI.  Formally, the assumptions of NIM and locality play similar roles in the derivation of the respective inequalities \cite{Paz1993}.

Leggett and Garg initially proposed an rf-SQUID flux qubit as a promising system on which to test their inequalities \cite{Leggett1985}, a proposal which was later refined by Tesche \cite{Tesche1990} (see also \cite{Peres1988,Leggett1989}).  Twenty-five years later, the first measured violation of a LGI was announced by Palacious-Laloy and coworkers \cite{Palacios-Laloy2010}.
This experiment differed from the Leggett-Garg proposal in a number of respects --- the superconducting qubit \cite{You05,Wendin07,Clark08,Girvin08,You11,Xiang13} was of the transmon type \cite{Koch2007}, and the measurements were continuous weak-, rather than instantaneous projective-, measurements \cite{Ruskov2006} --- but, nevertheless, the essence of the tested inequalities was as in Leggett and Garg.
Palacios-Laloy \etal \cite{Palacios-Laloy2010} found that their qubit violated a LGI, albeit with a single data point, with the conclusion being that their system does not admit a realistic, non-invasively-measurable description.
Signalling the death of MR, one commentator wrote ``no moon there'' \cite{Mooij2010} in refutation of the macrorealist belief, often associated with  Einstein \cite{Mermin1985}, that ``...the moon is there, even if I don't look at it''.

The Palacios-Laloy experiment was followed in the literature by a large number of further LGI tests  and, within a few years, violations had been reported in a wide range of different physical systems such as
photons \cite{Goggin2011,Xu2011,Dressel2011,Suzuki2012},
defect centres in diamond  \cite{Waldherr2011,George2013},
nuclear magnetic resonance \cite{Athalye2011,Souza2011,Katiyar2013},
phosphorus impurities in silicon \cite{Knee2012}, and milli-meter scale Nd$^{3+}$:YVO$_4$ crystals \cite{Zhou2012}.  Tests of LGIs on superconducting devices have also recently been revisited \cite{Groen2013}.
Table~\ref{TAB:overview} gives an overview of the different experimental systems in which LGI tests have presently been made
\begin{table}[tb]
\begin{center}
  \begin{tabular}{|c|c|c|}
    \hline
    Physical system & Measurement & Reference \\
    \hline
    \hline
    superconducting qubit & CWM & Palacios-Laloy \cite{Palacios-Laloy2010}\\
    ~ & W/SW & Groen \cite{Groen2013}\\
    nitrogen-vacancy centre & STAT & Waldherr \cite{Waldherr2011}\\
    ~ & W & George \cite{George2013}\\
    nuclear magnetic resonance & P & Athalye \cite{Athalye2011}, Souza \cite{Souza2011} \\
    ~ & INM & Katiyar \cite{Katiyar2013} \\
    photons & W/SW &  Goggin \cite{Goggin2011}, Dressel \cite{Dressel2011}, Suzuki \cite{Suzuki2012}\\
      ~ & P & Xu \cite{Xu2011}\\  
    Nd$^{3+}$:YVO$_4$ crystal & STAT & Zhou \cite{Zhou2012}\\
    phosphorus impurities in silicon & INM & Knee \cite{Knee2012} \\
    \hline
  \end{tabular}
  \end{center}
  \caption{
  An overview of the different physical systems in which LGI tests have been made.  The abbreviations for measurement types employed are:
  P: projective;
  CWM: continuous weak measurement;
  W/SW: weak/semi-weak point measurements;
  INM: ideal negative measurement; and 
  STAT: ``stationarity''.  The first author's name and reference are listed in the final column.
  \label{TAB:overview}
  }
\end{table}

One would be hard pressed to call the subjects of these studies ``macroscopic''.
Indeed, even for the qubit of \citer{Palacios-Laloy2010}, which was macroscopic in size,
subsequent analysis \cite{Palacios-Laloy2010a} has shown that the actual states involved in the LGI violation are not actually macroscopically-distinct (see \secref{SEC:SCqubit}).
Nevertheless, violations of the LGIs in ``microscopic'' systems (where really, we should speak of {\em microscopic realism} or just {\em realism} being at test) are of interest for a number of reasons.
If we share Leggett and Garg's goal of pursuing genuine macroscopic coherence, then the current experiments may be seen as a vital step towards scaling up to macroscopic objects.  As we will see, there are a number of non-trivial aspects to the LGIs, as well as a number of pitfalls, that make their experimental study anything but straightforward, even for microscopic systems.  
For example, with the exception of Refs.~\cite{Knee2012,Katiyar2013}, all of the LGI tests conducted so far suffer from the ``clumsiness loophole'' \cite{Wilde2012} that LGI violations can be ascribed to the unwitting invasivity of the measurements, rather that the absence of a macroscopic-real, NIM description of the system.  Without addressing this loophole, a devout macrorealist can safely ignore the challenge to his/her world view posed by these experiments.
Ironing out difficulties such as these in microscopic systems will increase the chance of successful pursuit of the genuine, macroscopic quarry.

Moreover, LGIs for microscopic systems are interesting in their own right.  One reason for this is the intimate connection between violations of the LGIs and the behaviour of a system under measurement.  Thus, the exploration of different measurement strategies has been a central theme of current experiments.
Furthermore, whilst the objects of the current experimental studies have all been ``good qubits'' \cite{iuliaROP}, there a number of situations where it is not clear to what extent the system is behaving quantum-mechanically.  If one accepts that the alternative to classical probabilities is quantum mechanics, then the LGIs provide an indicator of the ``quantumness'' of a system \cite{Miranowicz2010}.
The use of LGIs as such an indicator is coming to be appreciated across a growing number of areas, such as quantum transport \cite{Lambert2010,Emary2012,Emary2012a}, opto-mechanical and nano-mechanical devices near the quantum ground state  \cite{clerk11,LambertNem}, and even in the light-harvesting apparatus of biological organisms \cite{Wilde2010,Cheming,Lambert12}.  The connection between the ability to perform quantum-computations and violations of the LGI has also been studied by a number of authors \cite{Brukner2004,Morikoshi2006,Zukowski2010}.

The value of the LGIs lies in providing quantitative criteria to adjudge the line between classical and quantum physics.   In particular, Kofler and Brukner \cite{Kofler2007,Kofler2008} have used LGIs as a tool to study the emergence of the classical world from the quantum under coarse-grained measurements.
LGIs,  independent of questions of macroscopicity, are also at the centre of discussion on the similarities and differences between spatial and temporal correlations in quantum mechanics \cite{Brukner2004,Marcovitch2011,Marcovitch2011a}.

The aim of this review is to provide an introduction to the LGIs and to discuss recent developments in the field.
In \secref{SEC:formal} we discuss formal aspects of the LGIs, including their derivation, their underlying assumptions, and extensions.
We discuss the quantum violations of LGIs for the example of a qubit in \secref{SEC:qubit}, as this forms the basis for understanding many of the experimental results.
Section \ref{SEC:weak} considers the LGIs with weak measurements.  Sections \ref{SEC:SCqubit} to \ref{SEC:optics} discuss the various LGI experiments in the areas of superconducting qubits, nuclear spins, light-matter interactions and pure optics.
Sections \ref{SEC:QT} to \ref{SEC:NEMS} discuss theoretical proposals in the areas of quantum transport, photosynthesis and nanomechanical systems. In  \secref{SEC:related} we consider  constructions related to the LGIs, before concluding in \secref{SEC:concs}.

\section{Formalism \label{SEC:formal}}
We begin this section by first discussing the assumptions behind the LGIs and their implications.  We then give an explicit proof of \eq{K3intro} and then put this inequality in the context of a broad family of LGIs. Finally we discuss stationarity, ``entanglement-in-time'' and the entropic versions of the LGIs.

\subsection{Assumptions \label{SEC:ass}}

A crucial element of Leggett and Garg's work is the codification of how ``most physicists'' intuitively expect macroscopic objects to behave into a small set of principles or assumptions. Quoting directly from \citer{Leggett1985}, these principles read:
\begin{enumerate}
  \item[\A{1}] Macroscopic realism: A macroscopic system with two or more macroscopically distinct states available to it will at all times {\em be} in one or the other of these states;
  \item[\A{2}] Noninvasive measurability at the macroscopic level: It is possible, in principle, to determine the state of the system with arbitrarily small perturbation on its subsequent dynamics.
\end{enumerate}
In more-recent statements of the Leggett-Garg scheme \cite{Leggett2008,Kofler2008,Kofler2013}, a third assumption is often made explicit:
\begin{enumerate}
  \item[\A{3}] Induction: The outcome of a measurement on the system cannot be affected by what  will or will not be measured on it later.
\end{enumerate}
The conjunction of these properties has been called ``classicity''  \cite{Benatti1995} 
or, somewhat confusingly, ``macrorealism in the broader sense''  with assumption \A{1} in particular denoted ``macroscopic realism {\em per se}'' \cite{Leggett2002,Leggett2008,Kofler2007}.  We shall largely eschew these terms and refer to the assumptions explicitly to avoid confusion.
Under theories obeying \A{1-3}, Schr\"odinger's cat is, at each instant of time, either dead or alive, and which of these possibilities actually pertains can be divined through measurements that neither affect nor are influenced by its future history.  Assumptions \A{1-3} are thus in tune with our intuition about classical objects, but conflict strongly with quantum mechanics.

Whilst the derivation of the LGIs certainly relies on assumption \A{3}, so does much of our understanding of the natural world.  As this assumption reflects such basic notions about causality and the arrow of time, it has remained unchallenged in discussions of the source of LGI violation (but see \citer{Leggett2008} for a word of caution on this point).

Concerning assumption \A{1}, Peres notes \cite{Peres1988} that realism has ``at least as many definitions as there are authors'' and we will not attempt to give an account of this topic here (see rather \citer{Redhead1987}). 
The above definition of MR relies on the notion of  ``macroscopically distinct'' states.  A number of criteria exist by which this may be judged (see \citer{Nimmrichter2013} and references therein)
but we will defer a discussion of this point to later when we consider specific examples  (\secref{SEC:SCqubit} and \secref{SEC:crystal}).  
An important point, made by Maroney \cite{Maroney2012} and discussed in \secref{SEC:proof}, is that ``macroscopicity'' is not actually necessary for the derivation of the LGIs --- that the theory is ontic (i.e.,  realistic) is sufficient (in conjunction with \A{2} and \A{3}).

Whilst we can rely somewhat on our intuitive understanding of these two assumptions, assumption \A{2}, that of NIM, is more involved and has been the source of much discussion \cite{Ballentine1987,Leggett1987,Peres1988,Leggett1989, Tesche1990,Elby1992,Benatti1995,Leggett2008,Wilde2012}.
By way of clarification, let us first note that \A{2} presupposes \A{1}, in that a measurement is supposed to reveal a pre-existing property of a MR system.
Assumption \A{2}, therefore, defines a non-invasive measurement as one that would leave the state of the system unchanged by the measurement under a {\em macroscopic real} understanding of the system.  This clarification is important because a measurement on a quantum system can be ``non-invasive'' in the sense of \A{2}, i.e. a macrorealist might agree that the measurement could not disturb the system, and yet still be invasive in actuality because it causes a collapse of the system's wavefunction (a concept obviously absent from a macroscopic real description).  The statement of NIM for a quantum system is therefore counterfactual --- it refers to a property the system would have, if it were macroscopic real, which it is not.

Leggett and Garg \cite{Leggett1985} discuss how ``ideal negative measurements'' provide a method to probe a system in this non-invasive way.  
Consider that we are interested in the macroscopic variable $Q=\pm1$ and we can arrange it so that the detector only interacts with the system when it is in a state corresponding to $Q=+1$.  In this case, the absence of a detector response, combined with MR, allows us to infer the state of the system ($Q=-1$) even though our detector has not interacted with it.  
Provided that we only take such negative results into account, our measurement will be non-invasive in the sense of \A{2}, as the only results kept are those in which system and measuring apparatus did not interact.  Despite this, a quantum system can clearly still be affected by these measurements, since an \inm still induces wave function collapse \cite{Dicke1981}.

There are two distinct issues associated with the NIM assumption. 
Firstly, assuming that we can construct a measurement scheme to satisfy a macrorealist of its non-invasive credentials, then, setting \A{3} aside, a measured violation of a LGI implies either that MR must be rejected, {\em or} that it is intrinsically impossible to measure the system without disturbing its behaviour (or indeed both, as in quantum mechanics). 
Leggett writes \cite{Leggett1985,Leggett1988,Leggett2008} that NIM is such a ``natural corollary'' of MR that it is hard to see how NIM can fail but MR stay intact. However natural this may be, there is nothing in the violation of a LGI to preclude the possibility that the system is MR and yet not NIM \cite{Benatti1995} (Bohm-de Broglie would be a theory in this class \cite{Bohm1952,Bohm1952a}).
However, since even an invalidation of this intrinsic-NIM shows that the system is acting beyond what we expect from macroscopic objects, it is perhaps a moot point whether it is MR or the intrinsic NIM that fails.
As an aside, we note that the inability to test just MR is unavoidable, since realism by itself is consistent with the predictions of quantum theory \cite{Benatti1995,Leggett2008}.
In the LGIs, realism is tested in conjunction with NIM, just as it is tested in conjunction with locality in the spatial Bell's inequalities.

The second and by far the more serious problem associated with \A{2} is that, when confronted with a violation of the LGI, a macrorealist can always claim that, despite the best efforts of the experimentalist, his/her measurements were influencing the behaviour of the system in some unexpected way.
This is the so-called {\em ``clumsiness loophole''} \cite{Wilde2012} and a devout macrorealist can always exploit this avenue to refute the implications of a measured LGI violation since it is impossible to conclusively demonstrate that a physical measurement is in fact non-invasive.
One might think this possible by measuring the system at time $t$, again at time $t+\delta t$ and then comparing the results in the limit $\delta t \to 0$  \cite{Leggett1988}. If the results always agree, it would be tempting to conclude that the measurements are non-invasive. The problem with this is that, although this approach can exclude that the measurement is directly influencing macro-variable $Q$, it can not rule out that some unknown hidden variables are being influenced by the measurement, which then go on to affect the future time evolution. By appealing to such hidden variables, a macro-realist can always sidestep a LGI violation \cite{Knee2012}.

In Bell's inequalities, the analogous loophole is the communication loophole \cite{Clauser1978}.  This loophole can, however, be readily closed by making sure that the two measurements are space-like separated, so that events at one detector cannot influence the second during the duration of the experiment \cite{Scheidl2010}. 
Whilst a secure external physical principle (special relativity) is used to close this Bell inequality loophole, no such cast-iron defence exists for the LGI. 
The best one can hope for is strategies, such as ideal negative measurement, that make the explanation of LGI violations in terms of experimental clumsiness so contrived as to be unacceptable.
In this direction, \citer{Wilde2012} formulated an improved Leggett-Garg protocol that allows the clumsiness loophole to be narrowed.  Introducing the concept of an ``adroit measurement'' as one which, when enacted between the measurement times of the LGI, does not, by itself, affect the measured values of the Leggett-Garg correlation functions, the authors show that a violation of their updated protocol means that either the system is non-macrorealistic, or that two or more adroit-measurements, each individually non-invasive,  have somehow conspired to disturb the system.  This collusion is less plausible than independent non-invasive measurements, and the size of the loophole is correspondingly reduced.
We note that a number of ``loophole-free'' Bell tests have been proposed \cite{Kwiat1994,Huelga1995a,Fry1995,Rosenfeld2009} (see also \cite{Santos1991,Barrett2002,Kent2005}). Whether loophole-free Leggett-Garg protocols can be constructed is an open question.

\subsection{Proof of the LGIs \label{SEC:proof}}

The correlation function $C_{ij}$ is obtained from the joint probability $P_{ij}(Q_i,Q_j)$
of obtaining the results  $Q_i = Q(t_i)$ and $Q_j = Q(t_j)$ from measurements at times $t_i$, $t_j$ as
\beq
  C_{ij} = \sum_{Q_i,Q_j = \pm 1} Q_i Q_j P_{ij}(Q_i,Q_j)
  \label{EQ:ccfn}
  .
\eeq
The subscripts on $P$ remind us of when the measurements were made.
Assumption \A{1} means that, since observable $Q$ has a well-defined value at all times, even when left unmeasured, the two-time probability can be obtained as the marginal of a three-time probability distribution: 
\beq
  P_{ij}(Q_i,Q_j) = \sum_{Q_k;k\ne i,j} P_{ij}(Q_3,Q_2,Q_1),
\eeq
where the measurement subscripts have carried through.
Under MR alone, the three probabilities $P_{21}(Q_3,Q_2,Q_1)$, $P_{32}(Q_3,Q_2,Q_1)$ and $P_{31}(Q_3,Q_2,Q_1)$ required in the construction of \eq{K3intro} are independent, since measurements at different times may affect the evolution differently. Making the NIM assumption, \A{2}, however, precludes this possibility and all three probability distribution functions become the same: $P_{ij}(Q_3,Q_2,Q_1)=P(Q_3,Q_2,Q_1)$.  
This means that not only is the macro-variable $Q$ left unaltered by the measurements, but so must be any relevant hidden microscopic variables (not explicitly displayed here) that affect the time evolution.
This single probability can then be used to calculate all three correlation functions:
\beq
  C_{21} &=& P(+,+,+) -  P(+,+,-) -  P(-,-,+) +  P(-,-,-)
  \\
  \nonumber &&
           - P(+,-,+) +  P(+,-,-) +  P(-,+,+) -  P(-,+,-)
  ;
  \nonumber\\
  C_{32} &=& P(+,+,+) +  P(+,+,-) +  P(-,-,+) +  P(-,-,-)
  \\
  \nonumber &&
           - P(+,-,+) -  P(+,-,-) -  P(-,+,+) -  P(-,+,-)
  ;
  \nonumber\\
  C_{31} &=& P(+,+,+) -  P(+,+,-) -  P(-,-,+) +  P(-,-,-)
  \\
  \nonumber &&
           + P(+,-,+) -  P(+,-,-) -  P(-,+,+) +  P(-,+,-)
  ,
\eeq
where we have used the shorthand $P(+,+,+) = P(+1,+1,+1)$, etc.
Simple addition and completeness,
$
  \sum_{Q_3,Q_2,Q_1} P(Q_3,Q_2,Q_1) = 1
$,
give
\beq
  K_3 = C_{21} + C_{32} -C_{31} = 1-4\left[ P(+,-,+) + P(-,+,-)\right]
  \label{K3probs}
  .
\eeq
The choice of 
$
  P(+,-,+) = P(-,+,-) = 0
$
gives a value of $K_3 = 1 $, which is the upper bound of \eq{K3intro}.  Setting  $P(+,-,+) + P(-,+,-) = 1$ yields the lower bound: $K_3 \ge -3 $.  It is interesting to note that \eq{K3probs} implies an explanation of violations of the LGI in terms of negative probabilities \cite{Dirac1942}, a perspective discussed in \citer{Calarco1999} and employed in interpreting the experiments of \citer{Suzuki2012}.


An alternative proof of the LGIs has been given in terms of hidden-variable theories, e.g. \cite{Kofler2013,Dressel2011,Maroney2012}.  
We shall describe this proof in terms of the ``ontic model'' framework \cite{Spekkens2005,Rudolph2006,Harrigan2007,Harrigan2010}, and follow its terminology --- rather than hidden variables we will speak of the {\em ontic state} of the system, the real state of the system ``out there'' from which all physical properties can be derived.
To calculate the correlation functions $C_{ij}$, we assume that our system is prepared with some probability distribution $\mu(\zeta)$ over ontic states $\zeta$.   Measurement at time $t_i$ is represented by the outcome function, $\xi_i(Q_i|\zeta)$, which gives the probability of outcome $Q_i$ given ontic state $\zeta$. 
The probability of disturbance of the ontic state $\zeta \to \zeta'$ by the measurement is given by $\gamma_i(\zeta'|Q_i,\zeta)$.  In this way the generic ontic description for the joint probability function of two measurements reads
\beq
  P(Q_i,Q_j) = 
  \int d \zeta' d \zeta \,
  \xi_j(Q_j|\zeta')
  \gamma_i(\zeta'|Q_i,\zeta)
  \xi_i(Q_i|\zeta)
  \mu(\zeta)
\eeq
Under the NIM assumption \A{2}, the disturbance function leaves the ontic state untouched, $\gamma_M(\zeta'|Q,\zeta) = \delta(\zeta' - \zeta)$, whence
\beq  
  P(Q_i,Q_j) = 
  \int d \zeta \,
  \xi_j(Q_j|\zeta)
  \xi_i(Q_i|\zeta)
  \mu(\zeta)
\eeq
Inserting this into \eq{EQ:ccfn}, we obtain
\beq
  \ew{Q_i Q_j} &=& 
  \int d \zeta \,
  \sum_{Q_i,Q_j=\pm 1}
  Q_i Q_j
  \xi_j(Q_j|\zeta)
  \xi_i(Q_i|\zeta)
  \mu(\zeta)
  \nonumber\\
  &=& \int d \zeta \,
  \mu(\zeta)
  \ew{Q_i}_\zeta \ew{Q_j}_\zeta
\eeq
where $\ew{\ldots}_\zeta$ represents an expectation value for a given ontic state $\zeta$.  In these terms, $K_3$ of \eq{K3intro} can be written
\beq
  K_3 = \int d \zeta \,
  \mu(\zeta)
   \left[\frac{}{}
     \ew{Q_2}_\zeta \ew{Q_1}_\zeta
     + \ew{Q_3}_\zeta \ew{Q_2}_\zeta 
     - \ew{Q_3}_\zeta\ew{Q_1}_\zeta
   \right]
   .
\eeq
Since the expectation value of $Q_i$ is bounded in magnitude by unity, the bounds on $K_3$ are once again seen to be $-3 \le K_3 \le 1$.

From this derivation it is apparent that the LGIs are valid for any ontic (i.e. realistic) NIM theory.  Maroney \cite{Maroney2012} points out that this class of theories is larger than that of macroscopic realism, for which the ontic state of the system at any time must be of the form
\beq
  \mu(\zeta) = \sum_k p_k \nu_k(\zeta)
  ,
\eeq
where $\nu_k(\zeta)$ is a distribution of states which all share macroscopic property $k$ with respect to the relevant measurement $M$ (i.~e., $\nu_k>0$ only if $\xi_M(k|\zeta)=1$ for measurement outcome $k$).

\subsection{A family of inequalities}

The inequality of \eq{K3intro} is just one LGI to be found in the literature.  The most frequently encountered inequalities concern the $n$-measurement Leggett-Garg strings \cite{Athalye2011}
\beq
  K_n = C_{21} +  C_{32} +  C_{43} + \ldots + C_{n(n-1)} - C_{n1}
  \label{LGntermK}
  .
\eeq
Under assumptions \A{1-3}, these quantities are bounded as:
\beq
  \begin{array}{cc}
    -n \le K_n \le n-2 & n\ge 3, ~\mathrm{odd}
    ;
    \\
    -(n-2) \le K_n \le n-2 & n\ge 4,~\mathrm{even}
    .
  \end{array}
  \label{LGntermbounds}
\eeq
For $n$ odd, only the upper bound is of interest (at least, it is with projective measurements; see, however \cite{Williams2008,Williams2009}). For $n$ even, both bounds are relevant.  For these bounds to hold, the variable $Q$ need not necessarily be dichotomic $Q=\pm1$, but it must be bounded $|Q|\le 1$ \cite{Paz1993,Ruskov2006}.

Various symmetry properties of the above inequalities can be taken advantage of to derive further inequalities.  Firstly, the inequalities still hold under redefinition of the measured observables (providing they still obey $|Q|\le 1$) independently at each time.  In particular, we can redefine $Q \to -Q$ at various times in $K_n$ \cite{Ruskov2006}.  At third-order, this procedure generates the inequality
\beq
  -3 \le K_3'\le 1
  ;\quad\quad
   K_3' \equiv - C_{21} - C_{32} - C_{31}
   \label{K3'}
  ,
\eeq
which is the three-time inequality found in \citer{Leggett1985}. Moving to higher orders, this procedure allows us to generate inequalities for quantities as in \eq{LGntermK} but with any odd number of minus signs (rather than just the one).  At fourth-order there is only one distinct sign assignment:
\beq
  -2 \le C_{21} +  C_{32} +  C_{43} - C_{41} \le 2 
  \label{LGI_CHSH}
  ,
\eeq
which is equivalent to the four-term inequality of \citer{Leggett1985}.  At order five, there are three possibilities
\beq
  -5 \le &C_{21} +  C_{32} +  C_{43} + C_{54} - C_{51} \le 3
  \nonumber\\
  -5 \le &C_{21} +  C_{32} -  C_{43} - C_{54} - C_{51} \le 3
  \nonumber\\
  -5 \le - &C_{21} -  C_{32} -  C_{43} - C_{54} - C_{51} \le 3
  .
\eeq
Further inequalities may also be generated by permutation of the time-indices.

Avis \etal \cite{Avis2010} have given a characterisation of the complete space of LGIs formed with two-point correlation functions in terms of the geometry of cut polytopes.  In this scheme, the above LGIs of order $n\ge 4$  are all {\em reducible}, in the sense that they may be obtained from combinations of  ``triangle inequalities``, i.e. $K_3$, $K_3'$ and their time-permuted cousins.  The multi-time LGIs of \citer{Barbieri2009} and \citer{Wilde2012} are also of this reducible type. 
As an example of a higher-order {\em irreducible} LGI, Avis \etal describe the five-time ``pentagon inequality'',
\beq
  \sum_{i\le i <j\le 5} C_{ji} + 2 \ge 0
  ,
\eeq
which can be violated even when all relevant triangle inequalities are satisfied.
Reducibility does not necessarily render the inequalities for $K_n$ with $n\ge 4$ uninteresting.  For example, \citer{Wilde2012} takes advantage of higher-order reducible LGIs to address the clumsiness loophole.  Different reducible inequalities are also affected differently by dephasing (see \secref{SEC:dephase}).

\subsection{Stationarity \label{SEC:stat}}

If the correlation functions $C_{ij}=\ew{Q(t_i)Q(t_j)}$ are stationary, i.e. functions only of the time difference: $\tau = t_i-t_j$, then the $n$-measurement upper-bound inequality of \eq{LGntermbounds} obtains the simple form
\beq
  (n-1)\ew{Q(\tau) Q } -	\ew{Q([n-1]\tau)Q}
  \le n-2
  \label{LGnterm_stat}
  ,
\eeq
and the experimental effort required to test each of these LGIs is reduced to the measurement of just two correlation functions.  Let us reinforce that \eq{LGnterm_stat} is derived under the same assumptions as the original LGIs, \A{1-3}, but with the additional assumption that the correlation functions are stationary, a property which can be experimentally verified.

Huelga and coworkers \cite{Huelga1995,Huelga1996,Huelga1997,Waldherr2011,Zhou2012} have also discussed the derivation of Leggett-Garg-style inequalities under what they call ``stationarity''.  A typical example is
\beq
  P\rb{n,2t|n,0} -  \left[P\rb{n,t|n,0}\right]^2 \ge 0
  \label{LGI_Huelga}
  ,
\eeq
where $P\rb{n,t|n,0}$ is the conditional probability that, given that the system is in MR state $n$ at time $t=0$, it will be found in the same state at later time $t$.  
It is argued, e.g.~\citer{Waldherr2011}, that this inequality can be derived without NIM, and that MR and ``stationarity'' are sufficient, although the meaning of ``stationarity'' is left slightly open to interpretation.
We find that to derive \eq{LGI_Huelga} without NIM, the full set of assumptions required is:
\begin{enumerate}
 \item macroscopic realism; 
 \item time-translational invariance of the probabilities: $P\rb{n,t+t_0|n,t_0} = P\rb{n,t|n,0}$ for arbitrary $t_0$;   
 \item that the system is {\em Markovian};
 \item that the system is {\em prepared} in state $n$ at time. $t=0$; 
\end{enumerate}
The Markov assumption allows the probability $P\rb{n,2t|n,0} $ to be decomposed according to Chapman-Kolmogorov rules \cite{vanKampenbook} as 
\beq
  P\rb{n,2t|n,0}  = \sum_k P\rb{n,2t|k,t} P\rb{k,t|n,0},
\eeq  
where the sum is over all possible (MR) states of the system at time $t$. 
With time-translational invariance, the $k=n$ term in the sum cancels with the second term in \eq{LGI_Huelga} to give a non-negative quantity as stated.  
This formulation avoids having to make the NIM assumption by explicitly preparing the system in state $n$ at time $t=0$ and utilizing the Markov property that the subsequent evolution of the system is independent of whether the system entered a given state through preparation or in the course of its dynamics.
If the probabilities $P\rb{n,t|n,0}$ are obtained by making two-time measurements on an evolving system, then NIM once again has to assumed for \eq{LGI_Huelga} to hold (and assumption (iv) above, but not (i-iii), may be dropped).

A number of other authors have derived Leggett-Garg-type inequalities using  assumptions that are essentially equivalent to the Markov approximation, e.g. \cite{Foster1991,Elby1992,Zukowski2010}.  The Markov approximation is clearly  stronger than NIM --- NIM requires only that the system has no memory of whether it has been measured or not, 
Markovianity requires amnesia of its entire history.
In practice, the Markov assumption is as elusive, if not more so, than NIM.  Stated fully, the assumption is that the system is Markovian under a MR understanding, which, for a quantum system is an untestable proposition.  Maybe this macroscopic-Markov assumption can be made plausible, as is done with NIM, but a discussion of this point is lacking in the literature.
The combination of Markovanity and time-translational invariance corresponds to being able to write down a Markovian master equation for the populations of the complete set of macroscopic states that are thought to describe the system.
Violations of \eq{LGI_Huelga} by quantum systems can therefore be understood in terms of rewriting the coherent evolution of a quantum system as a {\em non-Markovian} rate equation for these probabilities by ``tracing out'' the coherences from the  Liouville-von-Neumann equation \cite{Nakajima01121958,Zwanzig1960,Emary2011}.
Finally on this point, we note that, whereas violations of \eq{LGI_Huelga} may be explained as a break-down in the Markovianity of the system, this does not apply to the full LGIs, which are valid whether the evolution is Markovian or not.

\subsection{Entanglement in time \label{SEC:EiT}}

There exists another class of inequality which can lay equal claim to the epithet ``temporal Bell's inequalities''.  A representative member is the temporal CHSH inequality discussed in \citer{Brukner2004} and \citer{Fritz2010} (see also \citer{DeZela2007} for a hidden-variables treatment). There, in each run of the experiment, Alice makes her dichotomic ($\pm1$) measurement at time $t_1$, whilst Bob makes his measurement at time $t_2>t_1$. They each have two choices ($i=1,2$) of detector setting, such that they measure variables $A_i$ and $B_i$ for Alice and Bob, respectively.  Under the Leggett-Garg assumptions \A{1-3} (see also \citer{Lapiedra2006} for a derivation based on a ``joint reality'' assumption) and in direct analogy with the CHSH inequality \cite{Clauser1969}, we obtain
\beq
  |\ew{B_1 A_1}+\ew{B_1 A_2}+\ew{B_2 A_1}-\ew{B_2 A_2}| \le 2
  \label{temporalCHSH}
  .
\eeq
A qubit can violate \eq{temporalCHSH} up to the Cirel'son bound of $2\sqrt{2}$ \cite{Cirelson1980}.

Comparison of spatial and temporal Bell inequalities has led Brukner \etal \cite{Brukner2004} to consider the possibility of ``entanglement-in-time'' in analogy with the usual entanglement responsible for the violations of the spatial inequality. Whilst this analogy works to a point, it is not complete. For example, in the extension to multi-partite entanglement, spatial entanglement is known to be monogamous \cite{Koashi2004}, but the temporal version was found to be polygamous.
Marcovitch and Reznik \cite{Marcovitch2011,Marcovitch2011a} have extended the temporal-spatial analogy by providing a precise mapping between two-time spatial and temporal correlation functions for general measurements and time evolutions.  
Central to this is the Choi-Jamio\l kowski isomorphism\cite{Jamiolkowski1972,Chiribella2009} between the space of bipartite systems $\rho_{AB} \in H_A \otimes H_B$ and the set of evolutions from $H_A$ to $H_B$.
In Marcovitch and Reznik's scheme, the temporal correlations must be obtained through weak measurements (the qubit is a special case where projective and weak measurements give analytically the same result).  This ``structural unification'' leads to a number of insights into the temporal Bell inequalities, as it allows the transfer of known results from the spatial to the temporal domain.  Further work on unifying spatial and temporal correlations in quantum mechanics can be found in Refs.~\cite{Leifer2013,Oreshkov2012,Fitzsimons2013}

As originally stated, LGIs and inequalities such as \eq{temporalCHSH} describe two physically-distinct scenarios: the first involves a set of measurements of the same operator at $n\ge3$ different times;  \eq{temporalCHSH}, in contrast, considers just two times but with different operator choices at each.
Formally, there is little difference between the two \cite{Markiewicz2013}.  We can map the LGI of \eq{LGI_CHSH} on to \eq{temporalCHSH} by simply assuming time evolutions such that $Q(t_1) = B_2$, $Q(t_2) = A_1$, $Q(t_3) = B_1$, and $Q(t_4) = A_2$.  This is similar to the situation tested in many current experiments. For example, Goggin \etal \cite{Goggin2011} essentially test the inequality
\beq
  \ew{Q_2} + \ew{Q_2 Q_3} -\ew{Q_3} \le 1
  ,
\eeq
with, quantum mechanically, $\hat{Q}_2 =\hat{\sigma}_z$ and $\hat{Q}_3 = \hat{\sigma}_x$ and no time evolution in between. In Ref.~\cite{Goggin2011} this is portrayed as a LGI, but it could equally well be interpreted as the three-term variant of \eq{temporalCHSH}, namely (cf. \citer{Bell1964})
\beq
  \ew{B_2 A_2} + \ew{B_1 A_1}-\ew{B_1 A_2} \le 1
  ;\quad
  A_1 = B_2,
\eeq
with choices $ \hat{A}_1 = \hat{B}_2=\hat{\sigma}_z$, $\hat{A}_2=\hat{\sigma}_x$, and  $\hat{B}_1$ an operator that returns a value of $+1$ on the initial state $\ket{\sigma}$ (see \secref{SEC:optics}).
The danger of this path is that, without the temporal structure of the LGIs (or indeed \eq{temporalCHSH}), if we are just free to pick the operators $\hat{O}(t_i)$ (or $ \hat{A}_i$ and $\hat{B}_i$) as we please, then these inequalities essentially just become a test of the properties of hand-picked noncommuting observables.  
This is far from the spirit of the LGI --- if one knew how to define and measure non-commuting observables for a macroscopic system, there would be no question of macroscopic-coherence to answer.

\subsection{Entropic Leggett-Garg Inequalities}

The underlying assumption behind the bounds of both the Leggett-Garg and Bell inequalities is the existence, independent of measurement, of a joint probability distribution that can provide information on all relevant marginals.
Braunstein and Caves \cite{Braunstein1988,Braunstein1990} used this assumption to formulate a set of entropic Bell inequalities based on the Shannon and conditional entropies of probability distributions measured by spatially-separated parties .
This technique has been adapted by Morikoshi \cite{Morikoshi2006} (and recently revisited by Usha Devi \etal~\cite{Devi2013}) to the Leggett-Garg, or temporal, setting.

Let $P(q_j, q_i)$ be the joint probability that measurements at times $t_i$ and $t_j$ of observable $Q$ (not necessarily dichotomic) give the results $Q(t_i)= q_i$ and  $Q(t_j)= q_j$.
In terms of the conditional probability $ P(q_i|q_j) = P(q_j,q_i)/P(q_j)$, the conditional entropy reads
\beq
  H[Q(t_i)|Q(t_j)] \equiv  - \sum_{q_j, q_i} P(q_j, q_i)  \log_2 P(q_i|q_j),
\eeq
Using the chain rule for conditional entropies and the fact that entropy never increases under conditioning, Morikoshi \cite{Morikoshi2006} derived the $N$-measurement inequality
\beq
  \!\!\!\!\!\!\!\!\!\!\!\!\!\!\!\!\!\!\!
  H(Q(t_N),...,Q(t_0)) &\leq& 
  H[Q(t_N)|Q(t_{N-1})] 
  + 
  \ldots +
  H[Q(t_1)|Q(t_0)] + H[Q(t_0)]
  \label{Morikoshi}
  ,
\eeq
where the left-hand-side is the joint entropy. He goes on to employ this temporal entropic LGI in an investigation of the role of quantum coherence in Grover's algorithm.

From \eq{Morikoshi} one can derive the temporal analogues of the spatial entropic Bell's inequalities by noting that the information contained in a set of variables is never smaller than that in a subset of them.  This gives, for example \cite{Devi2013},
\beq
  \sum_{k=1}^N H[Q(t_k)|Q(t_{k-1})] - H[Q(t_N)|Q(t_0)]\geq 0.
\eeq
The $N=3$ version of this inequality was recently investigated experimentally in \citer{Katiyar2013}.  One advantage of these inequalities is that they are not restricted to bounded dichotomic operators (the standard LGIs can be made to work with such operators too, but this requires redefinitions and partitioning, and is not unique).

\section{LGI violations of a qubit\label{SEC:qubit}}
Most experimental tests of LGIs to-date have been performed on two-level systems or {\em qubits}\cite{iuliaROP}, the most elementary of quantum systems.  It is thus of interest to look in-depth at the violation of the LGIs for this system.  Although we  consider only a very specific two-level example, it has been shown \cite{Kofler2008} that every non-trivial quantum evolution, irrespective of the nature or size of the system, allows one to violate a LGI, given the ability to make projective measurements on the initial state.

\subsection{Maximum violations}
The classical correlation functions $C_{ij} = \ew{Q_i Q_j}$ have no unique quantum analogue, due to issues of operator ordering.
In discussing the measurement of $K_n$ for a quantum system, the meaning of the correlation function $C_{ij}$ must be specified.  Implicit in the original work of Leggett-Garg was that these quantities be obtained with projective measurements, in which case the correlation functions may expressed in the same way as in \eq{EQ:ccfn}.
As Fritz has shown \cite{Fritz2010}, the correlators so obtained are equal to the symmetrised combination:
\beq
  C_{ij} = \frac{1}{2}\ew{\left\{\hat{Q}_i,\hat{Q}_j\right\}}
  \label{Fritz_sym_C}
  .
\eeq
Parameterising the qubit operators as
$
  \hat{Q}_i = \mathbf{a}_i \cdot \hat{\svec}
$,
with $\hat{\svec}$ the vector of Pauli matrices and $ \mathbf{a}_i$ a unit vector, and using the identity
$
  \rb{\mathbf{a}_2 \cdot \hat{\svec}}\rb{\mathbf{a}_3 \cdot \hat{\svec}} =
  \mathbf{a}_2 \cdot\mathbf{a}_3~ \hat{\mathds{1}}
  + i \hat{\svec} \cdot \rb{\mathbf{a}_2  \times\mathbf{a}_3}
$,
we obtain
\beq
  \textstyle{\frac{1}{2}} \ew{\left\{\hat{Q}_{i},\hat{Q}_{j}\right\} }
  = \mathbf{a}_i\cdot \mathbf{a}_j \ew{\hat{\mathds{1} }}
  = \mathbf{a}_i\cdot \mathbf{a}_j
  ,
\eeq
independent of initial conditions.  The $n$th-order Leggett-Garg parameter then reads
$
  K_n =
  \sum_{m=1}^{n-1}  \mathbf{a}_{m+1}\cdot \mathbf{a}_m
  -
   \mathbf{a}_n\cdot \mathbf{a}_1
$.  Finally, defining $\theta_{m}$ as the angle between vectors $\mathbf{a}_{m}$ and $\mathbf{a}_{m+1}$, we obtain
\beq
  K_n =
  \sum_{m=1}^{n-1}  \cos \rb{\theta_{m}}
  -
   \cos \rb{\sum_{m=1}^{n-1} \theta_{m}}
   .
\eeq
This quantity is maximised by setting all angles $\theta_m = \pi/n$, such that the maximum value for a qubit is
\beq
  K_n^\mathrm{max} = n \cos\frac{\pi}{n}
  .
\eeq
For the first few values of $n$, this gives values of 
\beq
  K_3^\mathrm{max}={\textstyle\frac{3}{2}}
  ;\quad
  K_4^\mathrm{max}=2 \sqrt{2}
  ;\quad
  K_5^\mathrm{max}={\textstyle\frac{5}{4}}\rb{1+\sqrt{5}}
  ;\quad
  K_6^\mathrm{max}=3 \sqrt{3},
  \label{qubit_bounds}
\eeq 
and so forth.  
An analogous classical derivation posits a spin with components $a_i^\alpha$, $\alpha=x,y,z$, at time $t_i$, such that our correlation functions read $\ew{Q_i Q_j} = \sum_\alpha a_i^\alpha a_j^\alpha v_0^\alpha$, with $v_0$ the initial vector of the system which, without loss of generality, we choose in the $z$ direction.  Classically,  we have then
$
  K_n
  =
  \sum_{m=1}^{n-1}  a^z_{m+1} a^z_m
  -
  a^z_n a^z_1
$,
which differs from the quantum case in that it only includes $z$ components.  This quantity is maximised by setting $a^z_m = \pm 1$ and since at least one of the $n$ terms will be negative for any such assignment, the maximum classical value is $n-2$ as in \eq{LGntermbounds}.

Equation~(\ref{Fritz_sym_C}) holds not just for a qubit but also for a quantum system of arbitrary size, provided that the observable $\hat{Q}$ is obtained as the difference between two projection operators (one onto the subspace corresponding to $Q=+1$, and one onto the $Q=-1$ subspace)\cite{Fritz2010}.  From this it follows that the maximum quantum values in \eq{qubit_bounds} also apply to systems of arbitrary size, provided that they are measured in this fashion \cite{Budroni2013}. More general measurements (for example, one measures precisely the state of the $N$-level system and assigns $Q=\pm 1$ values to each of these $N$ states) may give maximum violations of the LGIs that exceed these values \cite{Budroni2014}.

Violations of the LGIs can be associated with the non-commutativity of the operator $\hat{Q}$ with itself at different times. With the above parameterisation, we have the commutation relation
\beq
  \left[
    \hat{Q}_i,\hat{Q}_j
  \right]
  =
  2 i \hat{\svec}\cdot(\mathbf{a}_i \times \mathbf{a}_j)
  .
\eeq
Assuming that the vectors $\mathbf{a}_i$ all lie in the $x$-$z$ plane with equal angles between them, $\theta_i=\theta$, the commutators between relevant operator pairs are
\beq
  \left[
    \hat{Q}_2,\hat{Q}_1
  \right]
  =
  \left[
    \hat{Q}_3,\hat{Q}_2
  \right]
  =
  2 i \hat{\sigma}_y \sin \theta
  ;
  \quad
  \mathrm{and}
  \quad
  \quad
  \left[
    \hat{Q}_3,\hat{Q}_1
  \right]
  =
  2 i {\sigma}_y \sin 2\theta
 .
\eeq
The points where these commutators simultaneously vanish are the points where violations of the LGIs ($K_3$ and $K_3'$ together) disappear. Furthermore, the sum of the magnitudes of these commutators is proportional to $ 2|\sin \theta|+|\sin 2\theta|$, which is maximised by setting $\theta_1=\theta_2=\pm \pi/3$. Thus the points where the commutators are simultaneously maximised are the points where the LGI violations are greatest.

\subsection{Time evolution \label{SEC:qubit_time}}

\begin{figure}[tb]
  \begin{center}
    \includegraphics[width=0.75\columnwidth,clip=true]{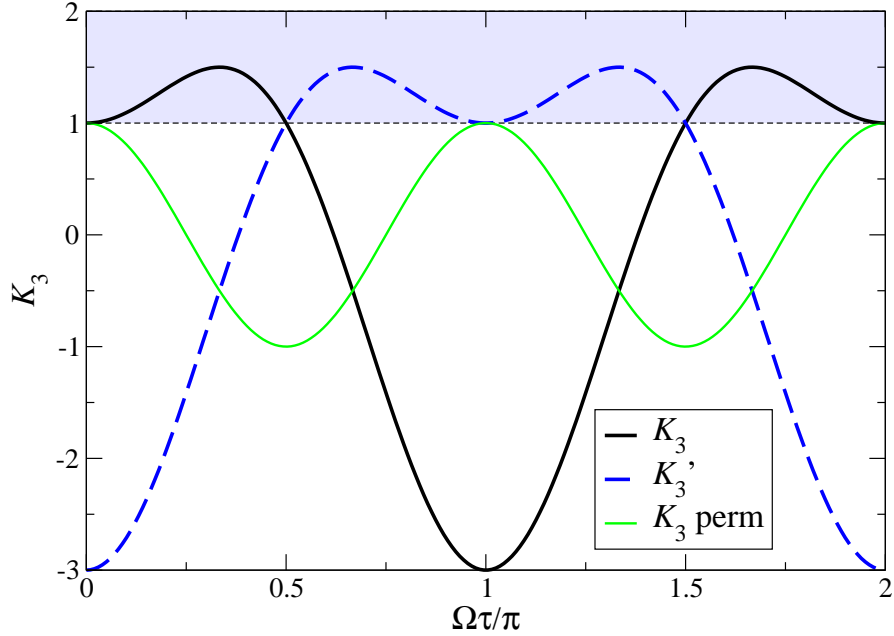}
  \end{center}
  \caption{
     (Color online)
     Third-order Leggett-Garg function $K_3$ with equi-spaced measurements for a qubit as a function of measurement-time spacing $\tau$.  The solid black curve shows the quantity $K_3$; the blue dashed curve the quantity  $K_3'$, and the thin green curve shows the function obtained by permuting the indices of $K_3$.  The blue shaded region denotes values of $K_3$ excluded by the Leggett-Garg inequality and thus incompatible with macroscopic realism and non-invasive measurability.
     A violation of one or the other of the $K_3$ and $K_3'$ inequalities occurs for all $\tau$ except at multiples of $\pi/2$.
    \label{FIG:K3qubit}
  }
\end{figure}

\begin{figure}[tb]
  \begin{center}
    \includegraphics[width=0.75\columnwidth,clip=true]{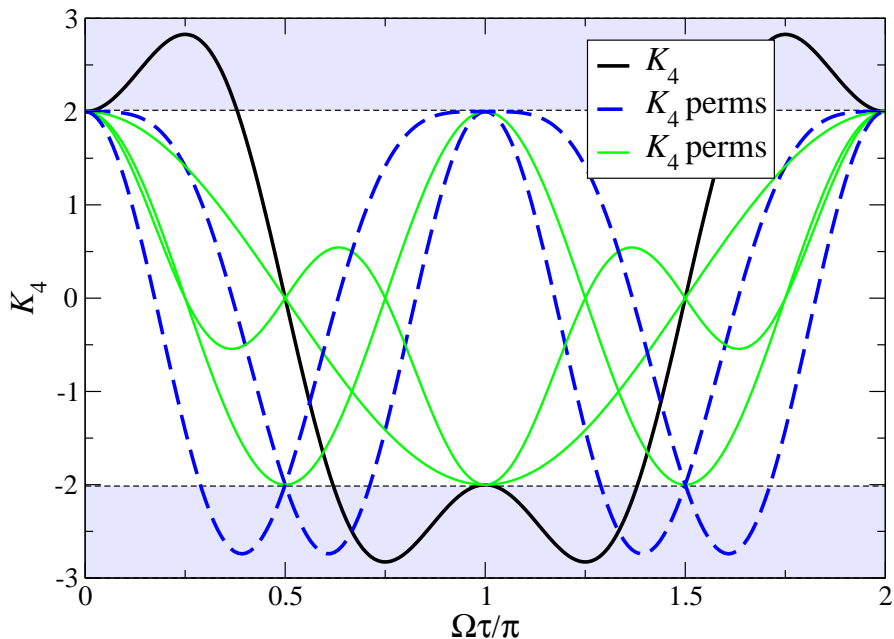}
  \end{center}
  \caption{
     (Color online)
     Same as in \fig{FIG:K3qubit},  but here for the fourth-order Leggett-Garg inequality, $|K_4|\le 2$.
     The thick black curve depicts $K_4$ itself; the other curves show $K_4$ with permutations of time indices.  Permutations that lead to violations of the inequality are plotted with blue dashed curves, and those that do not, with green thin curves.
    \label{FIG:K4qubit}
  }
\end{figure}

The canonical example of a time evolution that violates the LGIs is a qubit evolving under the Hamiltonian
\beq
  \hat{H}_\mathrm{qb} = \frac{1}{2} \Omega \hat{\sigma}_x
  ,
\eeq
and measured in the $z$-direction, $\hat{Q} = \hat{\sigma}_z$.  In this case, the correlation functions read \cite{Leggett1985}
\beq
  C_{ij} = \cos\Omega(t_i-t_j)
  ,
\eeq
and choosing equal time intervals, $t_{m+1} - t_m = \tau$, we obtain \cite{Athalye2011}
\beq
  K_n = (n-1) \cos\Omega \tau - \cos(n-1)\Omega \tau
  .
  \label{QB:Kntevol}
\eeq
The third-order $K_3$ is plotted in \fig{FIG:K3qubit}.  It oscillates as a function of the measurement time $\tau$ with maximum value of $3/2$ occurring at times $\Omega\tau = \pm\frac{\pi}{3}+2\pi k$, with $k$ an integer.  Only for certain ranges of $\tau$ is $K_3>1$.  At third order, permutation of the time indices only either recovers the original inequality, \eq{QB:Kntevol}, or generates the trivially-satisfied $\cos 2\Omega \tau < 1$.
The $ K_3'$ inequality of \eq{K3'}, however, yields the distinct
\beq
  -3 \le - 2 \cos\rb{\Omega \tau} - \cos\rb{2\Omega \tau} \le 1
  .
\eeq
$K_3'$ has maxima of $3/2$ at $\Omega\tau = \pm \frac{2\pi}{3} + 2\pi k$ and, as \fig{FIG:K3qubit} shows, is complementary to $K_3$ in that the violations of $K_3'$ fill in the gaps between those of $K_3$ \cite{Huelga1995}.
The only times for which no violation occurs
is when $\Omega \tau = \frac{k}{2} \pi$, where the system state is an eigenstate of the measurement operator and a QND measurement is performed \cite{Calarco1997}.

Turning now to the fourth-order inequality, from \eq{QB:Kntevol} we have
\beq
  K_4 = 3\cos\Omega \tau - \cos 3\Omega \tau,
\eeq
bounded from above and below by $\pm2$.  Of the $4!$ possible permutations of the time indices at fourth order, 6 distinct LGIs arise.
These results are plotted as a function of measurement time in \fig{FIG:K4qubit}.  For the qubit evolution considered here, only three of these permutations violate a LGI.
As in the $K_3$ case, at least one of the inequalities is violated for all values of $\Omega \tau$, except for multiples of $\pi/2$ . Note that for the even-order inequalities, both the upper and lower bounds are relevant.

Montina \cite{Montina2012} has shown that this pattern of violations of the LGIs for the qubit can be reproduced by a minimal classical model consisting of just four states --- the two states measured in the LGI test plus one ancillary bit --- combined with invasive measurements.

\subsection{Dephasing \label{SEC:dephase}}
The foregoing assumes unitary dynamics of the qubit. Contact with an environment can, however, induce dephasing,  the effects of which can be seen in, e.g., \fig{FIG:Palacios-Laloy}, where the oscillations of the Leggett-Garg parameter are damped with time.
From the perspective of obtaining the largest violations, the $K_3$-test is preferable to the $K_3'$-test because the maximum violation occurs at an earlier time with $K_3$, such that the effects of dephasing will be less.

A general framework for understanding the influence of non-unitary evolution on the maximal violations of the LGIs was given in \citer{Emary2013a}.  There  it was assumed that the observables $Q_\alpha(t_\alpha)$ could be chosen arbitrarily and independently at the three measurement times.  By maximising over all possible choices of these operators, the maximal possible violation for a given environment can be obtained.
This approach has an analogy with the treatment of the spatial Bell's inequalities, where maximization over measurement angles connects the value of the Bell correlator with a property of the input state, entanglement.   For the LGI, maximization over measurement angles reveals the connection between the Leggett-Garg correlator, $K$, and the non-unitary parameters of the dynamics.
A broad class of environments acting on a qubit was studied in \citer{Emary2013a}, modelled by generic quantum channels \cite{King2001,Keyl2002} acting in between measurements. For example, consider a depolarizing channel that serves to isotropically contract the Bloch sphere by a factor $-1\le c \le 1$ in each of the evolution periods, $t_1$ to $t_2$ and  $t_2$ to $t_3$.  The maximal value of $K_3$ in this case was found to be
\beq
  K_3^\mathrm{max} =
  \left\{
  \begin{array}{cc}
    |c|(1-|c|) & |c|\le 1/2 \\
    \frac{1}{2} + c^2 & |c|> 1/2
  \end{array}
  \right.
  \label{Kmaxdepolar}
  .
\eeq
Violations of the LGI are thus only possible when
$
  |c| > \frac{1}{\sqrt{2}}
$ and the violation thus shows a threshold behaviour --- if dephasing is too strong no Leggett-Garg violation can occur. This behaviour is not restricted to this particular example, but rather a general feature of unital (i.e., dephasing without relaxation) \cite{King2001} evolutions.

\section{LGIs and weak measurements\label{SEC:weak}}

In contrast to projective ones, weak measurements do not completely distinguish between possible values of the property being measured  \cite{Aharonov1988,Ashhab09,Ashhab2009a,Ashhab09b,Kofman2012}.  This ambiguity means that less information is gained about the system per experimental run and, quantum-mechanically, it means that such measurements may only partially collapse the system wavefunction. A true weak measurement is obtained in the limit of maximal  ambiguity and vanishing effect on the wavefunction.  We follow \citer{Dressel2011} in referring to measurements intermediate between weak and  projective as ``semi-weak''.

A number of works have derived \cite{Jordan2006,Ruskov2006,Williams2008} and tested \cite{Palacios-Laloy2010,Goggin2011,Dressel2011,Suzuki2012}
Leggett-Garg-like inequalities with semi-weak measurements and it is the aim of this section to elucidate how these tests differ from the standard LGI tests and from each other.
%

\subsection{Weakness and ambiguity}

As emphasised in \citer{Dressel2012}, weak measurements can be introduced  classically through the notion of an {\em ambiguous} detector.  Let us assume that we measure a system with a detector that gives response $q$ (assumed continuous here, but this need not be) to system variable $Q = \pm 1$ with probability $P(q|Q)$.   One can arrange that this ambiguous detector is calibrated such that the ambiguously-measured ensemble average
\beq
  \ew{q} \equiv \sum_Q \int dq \; q \, P(q|Q) \, P(Q)
  ,
\eeq
with $P(Q)$ the distribution of system variable is the same as would be measured with an unambiguous one, namely $\ew{Q} = \sum_Q P(Q) Q $. With this constraint the range of possible values of $q$ will exceed the original range of system variable $Q$.  

Quantum-mechanically, this situation can be expressed in terms of Kraus operators \cite{Kraus1971,Kraus1983}, where a single instantaneous semi-weak measurement of observable $\hat{Q}$ yields a result $q$ and changes the state of the system as
\beq
  \hat{\rho} \, \to\, \hat{\rho}_1(q) \,=\, 
  \hat{\mathcal{K}}(q) \, \hat{\rho}\, \hat{\mathcal{K}}^\dag(q)
  \label{rhoKraus}
  ,
\eeq
with Kraus operator $\hat{\mathcal{K}}(q)$. The probability of obtaining  outcome $q$ is
$
 P (q) = \mathrm{Tr} \hat{\rho}_1(q)
$.
As example, let us consider that $q$ is Gaussian-distributed about the eigenvalues of $\hat{Q}$ with a Kraus operator of the form \cite{Bednorz2012}
\beq
  \hat{\mathcal{K}}(q) = \rb{2\lambda/\pi}^{1/4} \exp\left[-\lambda(q-\hat{Q})^2\right]
  \label{Kraus}
.
\eeq
Here, the parameter $\lambda \ge0$ characterises the strength of the measurement; for $\lambda \to \infty$ we obtain a strong, projective measurement (corresponding to an unambiguous classical measurement), whereas $\lambda \to 0$ corresponds to the weak-measurement limit (corresponding to maximum ambiguity). 
The Kraus operator is defined such that the expectation $\ew{q} \equiv \int dq q P(q)$ is, as above, consistent with that of a projective measurement, $\ew{\hat{Q}} = \mathrm{Tr} \hat{Q}\hat{\rho}$.
In the language of \citer{Dressel2010,Dressel2012}, the detector response $q$ is a ``contextual value'' in a generalized spectrum for $\hat{Q}$ that depends on the context of the specific detector being used.

\subsection{Two-point Leggett-Garg inequalities \label{SEC:2point}}
There are a number of different ways in which weak measurements could be or have been deployed in LGI tests.
Most obviously, we could replace the projective measurements of the standard LGI procedure by weak measurements.  To be clear, the standard procedure for obtaining $K_3$ involves making three different types of experimental run and measuring each of the three correlation functions $C_{ij}$ separately. We will refer to this way of obtaining the $C_{ij}$ as the ``two-point method'' since in any given run, the system is only measured at two points in time.
For the qubit of \secref{SEC:qubit_time}, exchanging projective measurements for semi-weak ones leaves the two-point correlation functions $C_{ij}$ entirely unaltered. This is consistent with the observation of Fritz \cite{Fritz2010} that the projectively-measured correlation functions are identical to those obtained in the weak-measurement limit, $\frac{1}{2} \ew{\left\{\hat{Q}_j,\hat{Q}_i\right\}}$ \cite{Wang2002}.  The question of LGI violations when measured in the two-point fashion is thus independent of measurement strength.  This holds for a qubit, but not necessarily for systems of larger dimension.  Indeed, Kofler and Brukner \cite{Kofler2007,Kofler2008} have shown that `fuzzy' measurements on a quantum system can explain the emergence of classical behaviour (in this case, the compliance with the LGI) as the size of the system increases.

\subsection{Three-point Leggett-Garg inequalities \label{SEC:3point}}

The second, and far more interesting, approach with weak measurements departs from the original Leggett-Garg protocol and constructs $K_3$ (we shall only discuss this simplest form) by measuring the system at all three times in each run.  We shall refer to this method of determining $K_3$ as the ``three-point method''.  
Conducting LGI tests with weak measurements in this manner was first proposed in \citer{Jordan2006}, albeit there the measurements were performed in a repeatedly-kicked fashion.
The authors of \citer{Jordan2006} refer to the inequalities so-defined as {\em generalised LGIs} to indicate that they are different in kind to those of the original LGI proposal.
Violations of this type of LGI have been probed in several recent experiments \cite{Goggin2011,Dressel2011,Suzuki2012}, to be discussed in \secref{SEC:optics_weak}.
We note that a similar reformulation of the spatial Bell's inequalities was given in \citer{Marcovitch2010}.

A proof that the inequality $K_3\le 1$ still holds when the measurements are ambiguous can be obtained with a slight adaption of the  proof given by Dressel~\etal~\cite{Dressel2011} for their two-party inequality. In terms of violating this three-point LGI, the strength of the first and last measurements is irrelevant \cite{Goggin2011}, so we shall only consider that the middle measurement is semi-weak.
Classically, repeated runs of the three-point experiment furnish us with the probabilities $P(Q_3,q_2,Q_1)$, where $Q_3$ and $Q_1$ are dichotomic system variables at times $t_3$ and $t_1$ obtained from unambiguous measurements, and $q_2$ is the output of our ambiguous detector set to measure system variable $Q_2$ at time $t_2$.  For simplicity, we shall assume that we prepare the system in the state $Q_1=+1$, such that the probability reads $P(Q_3,q_2,Q_1)=P(Q_3,q_2)\delta_{Q_1,+1}$.
The quantity $K_3$ constructed from these three-point probabilities is then
\beq
  K_3 = \sum_{Q_3}\int d q_2 \; P(Q_3,q_2) \rb{q_2 + Q_3 q_2 - Q_3}
  .
  \label{K3weak}
\eeq
If we were to make the measurement at $t_2$  unambiguous and $q_2$ is restricted to the values $\pm1$, it is clear that this quantity is bounded as $-3 \le K_3 \le 1$.

To determine the bounds on $K_3$ when the $q_2$-measurement is ambiguous, we may  modifiy the argument of \secref{SEC:proof} in terms of ontic states (hidden variables in \citer{Dressel2011}). Under the assumption of realism and NIM 
\eq{K3weak} can be written as 
\beq
  K_3 = \int d\zeta \mu(\zeta) 
   \left[\frac{}{}
     \ew{q_2}_\zeta 
     + \ew{Q_3}_\zeta \ew{q_2}_\zeta 
     - \ew{Q_3}_\zeta
   \right]
   .
\eeq 
Since the expectation value from an ambiguous detector is identical with that of the variable itself, the magnitude $|\ew{q_2}_\xi|$ is bounded by unity.  The bounds on $K_3$ measured in this way are thus identical to those when measured projectively, i.e.~$-3 \le K_3 \le 1$.
Thus, a violation of this three-point LGI means that the middle measurement must have been ambiguous and that one of the standard Leggett-Garg assumptions \A{1-3} breaks down for the system.
It is interesting to compare how the two-point and three-point inequalities admit violations.
In the two-point LGI, it is the incompatibility between the independently-assessed two-point correlation functions with a single three-point joint probability distribution function that is the source of the LGI violations.
In the three-point case, it is the fact that $q$ is not restricted to the range of the measured variable $Q$ that opens up the scope for $K_3$ to exceed unity in the first place.  This, coupled with the fact that the quantum measurement is invasive permits the violation.

As an example of this type of violation we can consider a qubit with parameters as in \secref{SEC:qubit}, initialised in the state $\ket{+}$ (corresponding to $Q_1 = +1$) and measured at time $t_2$ with a detector described by the Kraus operators of \eq{Kraus}. 
For equally-spaced measurement times (spacing $\tau$), the requisite probability may be obtained as
\beq
 P(Q_3, q_2) =  \left|\bra{Q_3}\hat U(\tau)\hat\mathcal{K}(q_2)\hat U(\tau)\ket{+}\right|^2
 ,
\eeq
with unitary time-evolutions operator $\hat U(\tau) = \exp\rb{-i \hat H \tau}$,
such that the $K_3$-parameter reads
\beq
  K_3 = 2 \cos \Omega \tau
  - \exp\rb{-\lambda}
  \rb{
    \cosh \lambda \cos 2 \Omega \tau - \sinh \lambda
  },
\eeq
with $\lambda$ the strength parameter of the middle measurement.  In the limit $\lambda \to \infty$, we obtain
\beq
  K_3 = 2 \cos \Omega \tau - \cos^2 \Omega \tau
  ,
\eeq which is always less than or equal to one. Thus with projective measurements, we recover the expected result that calculating marginals from the projectively-measured three-point distribution can not violate the LGI \cite{Ballentine1987,Leggett1987}.  However in the opposite limit, $\lambda \to 0$, the weakly-measured $K_3$ here becomes the same as that of the undamped qubit, \eq{QB:Kntevol}, with the same pattern of LGI violations.  Indeed, all non-infinite values of $\lambda$ permit  LGI violations.  
It may at first seem strange that maximum violations are obtained in the limit
$\lambda \to 0$ when in this limit, the influence of the measurement on the systems wavefunction is negligible. However, this must be understood as the result of limiting process where, in order to obtain reliable statistics, the number of runs of the experiment also diverges.

Thus, providing that the intermediate measurements are semi-weak to some degree, an $n$-term LGI based on marginals calculated from an $n$-point measurement can be violated.
We note that the probability function $P(Q_3,q_2)$ calculated here is a genuine probability in that it is normalised and non-negative. The LGI is violated because the marginals derived from $P(Q_3,q_2)$ are
inconsistent with the NIM assumption. A quasi-probability can be extracted from $P(Q_3,q_2)$ by subtracting the detector noise \cite{Bednorz2012}. This quasi-probability can be negative, underlining the quantum origins of the violations. Curiously, the correlation functions $C_{ij}$ are the same whether one calculates them using the full probability distribution or the quasi-probability equivalent. This is expected to be a property of the two-level system only.
%

\subsection{Leggett-Garg inequalities with continuous weak measurements}

The final type of weak measurement to be discussed here is the {\em continuous weak measurement}, which is important particularly in the solid state where the typical measurement device is permanently attached to the system \cite{Ashhab09,Ashhab09b}.
Such a continuous weak measurement can be described by extending the Kraus-operator approach above \cite{Bednorz2012}. However, violations of a LGI with continuous weak measurement were first discussed by Ruskov \etal \cite{Ruskov2006} within a quantum stochastic approach, and it is instructive to consider this presentation.

Rather than measuring the qubit observable $Q(t)$ directly, the continuous weak measurement detector obtains the noisy signal
\beq
  I(t) = I_0 + (\Delta I/2)  Q(t) + \xi(t)
  ,
\eeq
where $I_0$ is an offset, $\Delta I $ the signal response and $\xi(t)$ a stochastic variable representing Gaussian white noise with zero temporal average
\beq
  \ew{\xi(t)}_t \equiv \lim_{T \to \infty}
  \frac{1}{T}\int_{-T/2}^{T/2} \xi(t) dt =0
  ,
\eeq
and $\delta$-correlation
\beq
  \ew{\xi(t)\xi(t+\tau)}_t = \frac{1}{2}S_0 \delta(\tau),
\eeq
with spectral density $S_0$ \cite{Gardiner2004}.  The time-averaged correlation function of the detector variable
\beq
  C_I(\tau) =
  \ew{
  [I(t) - I_0][I(t+\tau) - I_0]
  }_t
  ,
\eeq
consists of four contributions. However, by specifying $\tau > 0$, the detector noise contribution is avoided and, provided that the qubit doesn't anticipate the future behaviour of the detector (cf. \A{3} ), the term $\ew{Q(t)\xi(t+\tau)}$ also vanishes.  The quantity  $\ew{\xi(t) Q(t+\tau)}$ describes the back-action of the detector upon the qubit.  In line with the NIM assumption of the projective LGI, we assume that the measurement set-up can be arranged such that this term vanishes.
Assumptions \A{2} and \A{3} are thus expressed in the continuous weak measurement case by the statement
\beq
  \ew{Q(t) \xi(t+\tau)}_t = \ew{\xi(t) Q(t+\tau)}_t=0
  \label{CWMA2A3}
  ,
\eeq
which is postulated to hold true for macroscopic systems.  Under these assumptions, we obtain a direct relation between the detector correlation function and that of the system:
\beq
  C_I(\tau) =
  (\Delta I/2)^2 \ew{Q(t)Q(t+\tau)}_t
  .
\eeq
We can then use the known LGIs for $Q$ to write down inequalities for the continuous weak measurement correlation functions, e.g.~\cite{Ruskov2006}
\beq
  C_I(\tau_1) + C_I(\tau_2) - C_I(\tau_1+\tau_2) \le (\Delta I/2)^2
  \label{LGIweak}
  .
\eeq
In this way of testing the LGIs, the averages are temporal averages, which has the practical advantage that a correlation function may be obtained in a single run and the theoretical advantage that any possible issues with ensembles \cite{Leggett2002} are avoided.

Ruskov {\em et al.} \cite{Ruskov2006} calculated these correlation functions for a double-quantum-dot charge qubit coupled to a quantum-point-contact detector \cite{Korotkov1999,Korotkov2001,Korotkov2001a}.  With qubit Hamiltonian and measurement operator as in \secref{SEC:qubit_time} they found
\beq
  C_I(\tau)  =\rb{\frac{\Delta I}{2}}^2 \exp\rb{-\Gamma \tau/2}
  \rb{
    \cos \tilde{\Omega}\tau
    + \frac{\Gamma}{2\tilde \Omega}\sin \tilde\Omega \tau
  }
  \label{CIweak}
  ,
\eeq
with shifted frequency
$
  \tilde \Omega \equiv \sqrt{\Omega^2 - \Gamma^2}
$
and total dephasing rate $\Gamma = \gamma + (\Delta I)^2/4S_0$ that includes an environmental  contribution, $\gamma$, and one arising from the coupling to the detector.  In the limit of weak system-detector coupling and good isolation from the environment, $\Gamma/\Omega \ll 1$, the ratio $C_I(\tau) / (\Delta I /2)^2$ recovers the correlation functions of \secref{SEC:qubit_time} and the corresponding pattern of LGI violations result.
This continuous weak measurement formalism was utilised in the Palacios-Laloy experiment \cite{Palacios-Laloy2010}, which we discuss in \secref{SEC:SCqubit}.

\subsection{Weak versus non-invasive measurements \label{weakvsNIM}}

It is important to stress that a weak measurement is not necessarily a non-invasive (in the sense of \A{2}) one.
Whereas the strength/weakness of a measurement relates to the degree of ambiguity in the results, the non-invasivity of \A{2} is the property that the measurement should not influence the future time evolution of a macroscopic-real system. With this distinction made, it is obvious that an ambiguous measurement performed clumsily can be just as invasive as an unambiguous one.

Where the confusion arises is that, from a purely quantum-mechanical perspective, a (strictly) weak measurement induces a vanishing degree of wavefunction collapse, thus minimising the ``quantum-mechanical invasiveness'' per run.  This invasiveness, though, is purely a quantum-mechanical in origin and has no meaning for a macrorealist.  It therefore can not enter into his/her opinion on whether a system is being measured invasively or not.

It might be argued that weakness of the measurement arises from a weak physical coupling between the system and detector and therefore any effects of the detector must be minimal.  However, this isn't necessarily the case --- a weak measurement can be performed with a strongly coupled detector, provided that the detector is very noisy.
Moreover, concentrating on the continuous weak measurement case,  \eq{LGIweak} shows that the threshold for LGI violations is $(\Delta I/2)^2$. Thus, an inadvertent invasive component of the measurement need only have a coupling strength similar to the system-detector coupling to exert an influence on whether a LGI is violated or not.  The clumsiness problem remains, no matter how weak the coupling is made.

The NIM criteria in the continuous weak measurement case is actually very clear, and it is distinct from notions of weakness --- to claim NIM, one has to be able to convince a macrorealist that \eq{CWMA2A3} holds \cite{Wilde2012}.  
Of course, for a quantum-mechanical system, the cross-correlator $\ew{\xi(t) Q(t+\tau)}_t$ will not be zero (except for the singular and uninteresting case of QND measurements) due to the unavoidable collapse-related back-action of the detector on the system.  This back-action is precisely the reason why \eq{LGIweak} can be violated in the quantum-mechanical case \cite{Ruskov2006,Wilde2012}. 
Thus, how to counterfactually assert that the measurement is non-invasive is just as much of a problem with weak measurements as it is with strong ones.  This problem seems to have gone unaddressed in the literature.

\section{Superconducting qubits and the first experimental violation of a LGI \label{SEC:SCqubit}}

The first experimental test of a LGI came not with the rf-SQUID of the original Leggett-Garg proposal but rather a superconducting charge qubit of transmon type formed by a Cooper pair box shunted by a microwave transmission line \cite{You05,You11,Xiang13,Koch2007,Schreier2008}.  Due to the large ratio of Josephson- to charging- energy, such qubits show a reduced sensitivity to charge noise, making them good candidates for observing quantum coherent phenomena.

\begin{figure}[tb]
  \begin{center}
    \includegraphics[width=0.7\columnwidth,clip=true]{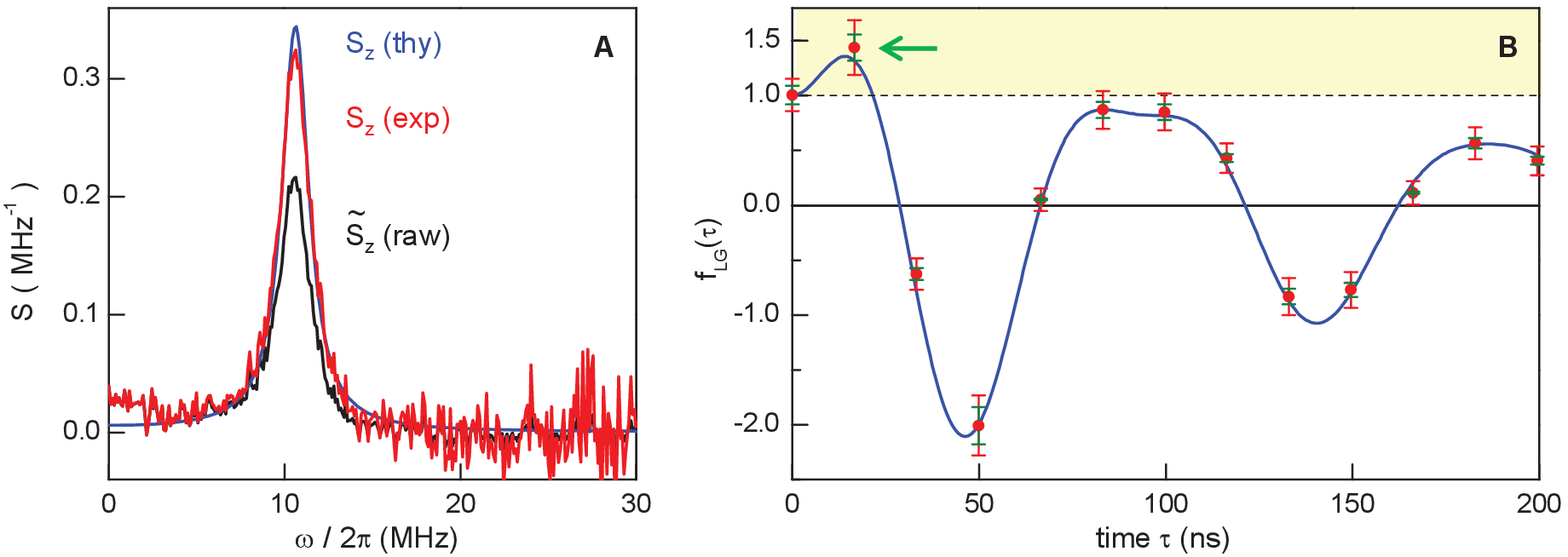}  
  \end{center}
  \caption{
     (Colour online) Experimental results from the measurement of the three-term continuous-weak-measurement Leggett-Garg inequality, \eq{EQ:fLGweak}, for a superconducting transmon qubit.
     Red points are experimental data points;  the blue line, theoretical quantum prediction and
     yellow, the region forbidden under Leggett-Garg assumptions. The data point marked with the arrow indicates a violation at short times.
     Figure from \citer{Palacios-Laloy2010}.
    \label{FIG:Palacios-Laloy}
  }
\end{figure}

In the circuit-QED experiments of Palacios-Laloy~{\em et al.}~\cite{Palacios-Laloy2010} the qubit was both driven and measured by a microwave resonator with the measurement in the continuous-weak-measurement paradigm discussed previously.  Under MR assumptions about the response of the microwave resonator and the subtraction of detector noise, the correlation functions $C_I(\tau)$ were extracted from the measured spectral density of the resonator. From these, the weakly-measured LGI 
\beq
  f_{LG}(\tau) = 2C_I(\tau) - C_I(\tau) \le 1
  \label{EQ:fLGweak}
  ,
\eeq
was tested.  The experimental results are reproduced in \fig{FIG:Palacios-Laloy} and good agreement with the quantum-mechanical predications was observed.  A violation of \eq{EQ:fLGweak} was observed, but only as a single data point with
$f_\mathrm{LG}(\tau) = 1.37 \pm 0.13$ at $\tau \sim \pi/3\omega_R$, with $\omega_R$ the Rabi frequency of the qubit.

Palacios-Laloy \cite{Palacios-Laloy2010a} gives an interesting discussion of whether their experiment should be seen as a test of macroscopic coherence and concludes that `` [the] experiment does not involve superpositions of macroscopic
states but rather superpositions of microscopically-distinct states of a macroscopic body''. This conclusion is based on two criteria for macroscopic distinctness of two states, set forth by Leggett \cite{Leggett1980,Leggett2002}:
\begin{itemize}
  \item The extensive difference, $\mathfrak{L}$, is the difference between the  expectation values of the measured observable between the two states (e.g., the magnetic flux), scaled to some relevant atomic reference unit (e.g., the flux quantum);
  \item  The disconnectivity $\mathfrak{D}$ is a measure of the type of entanglement of the state: a density matrix with irreducible $M$-body correlations has a disconnectivity $\mathfrak{D}=M$.
\end{itemize}
In terms of these measures, macroscopic coherence implies $\mathfrak{L}\sim\mathfrak{D} \sim N$, with $N$ the number of microscopic constituents of the macropscopic body.  This was found to be the case by Leggett \cite{Leggett2002} for the rf-SQUID of \citer{Leggett1985},  although this was later found to be overly optimistic \cite{Korsbakken2009}.  In contrast, for the transmon qubit, 
Palacios-Laloy report a value of $\mathfrak{L}\sim10^{-7}$, since the difference in flux of the two states is small, and a disconnectivity of $\mathfrak{D} = 2$, since the Cooper-pair box can be described purely in terms of two-body wavefunctions.  Thus, although certain aspects of the experiment are macroscopic (e.g., the actual physical dimensions of the system), the superposition states involved in the LGI violations are only microscopic.
No justification of the non-invasivity of the measurement assumption was made.

In a recent experiment, Groen \etal \cite{Groen2013} have also realised a measurement of LGIs in a transmon qubit, but this time using a second transmon qubit as read-out device.  The set-up allowed the strength of the measurement to be controlled and the relationship between weak values and LGIs to be investigated. These themes are taken up in different contexts in \secref{SEC:optics_weak}.

\section{Nuclear spins\label{SEC:nuke}}

A number of groups have reported experimental tests of LGIs with nuclear-spin qubits.
Waldherr~\etal~\cite{Waldherr2011} studied a nuclear spin at a nitrogen-vacancy defect in diamond undergoing Rabi oscillations induced by rf pulses.  The state of the nuclear spin was read-out by a defect electron and Huelga's inequality of \eq{LGI_Huelga} was tested.
George~\etal~\cite{George2013} also performed experiments with an NV centre, but considered the nuclear spin as a three-level system and investigated the relation between quantum strategies in the ``three-box'' quantum game \cite{Aharonov2001,Maroney2012} and violations of LGIs.
Several experimental tests of LGIs in liquid-state (room temperature) NMR systems have been reported \cite{Souza2011,Athalye2011,Katiyar2013}. In all cases, the experiments were conducted on the chloroform molecule in which spin-half Carbon-13 nuclei were probed using the spin of Hydrogen-1 nuclei. Souza~\etal~\cite{Souza2011} considered the $K_3$ inequality whereas  Athlaye~\etal~\cite{Athalye2011} considered both $K_3$ and $K_4$ inequalities.  Katiyar~\etal~\cite{Katiyar2013} investigated an entropic LGI and also compared marginals obtained from the directly-measured three-point joint probability distribution $P(Q_3,Q_2,Q_1)$ with those from the two-point LGI measurements and found the mismatch responsible for LGI violations in a quantum system.
The interpretation of the measurements of Souza~{\em et al.}~\cite{Souza2011} as constituting a meaningful violation of a LGI has been criticised \cite{Knee2012a} (see also \citer{Souza2012}) and some of these objections apply more generally to other NMR tests of the LGIs (see later in this section).
Finally, Knee \etal \cite{Knee2012} considered spin-bearing phosphorus impurities in silicon with a nuclear spin as system qubit and an electron spin as an ancillary read-out qubit.

\begin{figure}[tb]
  \begin{center}
    \includegraphics[width=0.65\columnwidth,clip=true]{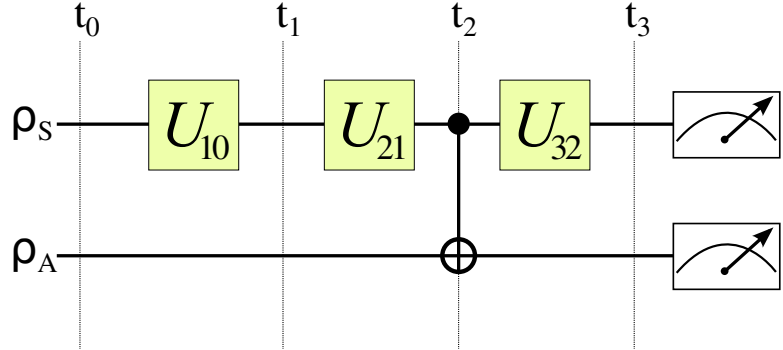}
  \end{center}
  \caption{
    (Color online)
    Quantum circuit to non-invasively measure part of the correlation function $C_{32}$.
    The system qubit is prepared in state $\rho_S$ and the ancilla qubit in $\rho_A$.  The time evolution of the system is induced by the unitary operators $U_{ji}$ acting between times $t_i$ and $t_j$.
    The measurement at $t_2$ is carried out with a CNOT-gate, in which the state of the ancilla is flipped if the system is in the $\uparrow$-state and left unaltered if the system is in the $\downarrow$ state.  The ancilla is then read-out at the end and only results where no flip has occurred are kept.
    The measurement of the system at $t_3$ can be performed invasively.
    This circuit is then repeated with an anti-CNOT gate instead and the measurements combined to build $C_{32}$ from ideal negative measurements.
    Adapted from Knee \etal \cite{Knee2012}.
    \label{FIG:Knee}
  }
\end{figure}

All of these works make use of a probe or {\em ancilla} qubit to perform the measurement \cite{Paz1993}.  In \citer{Knee2012}, this technique was used to realise an \inm and we will discuss this method here.
A quantum circuit for the measurement of the correlation functions $C_{ij}$ is shown in \fig{FIG:Knee}.
The essential ingredient is a CNOT gate acting on the system-ancilla pair with the system qubit as control and ancilla as target \cite{Nielsen2000}.  The CNOT gate performs a bit-flip of the ancilla if, for instance, the control is in state $\downarrow$, but leaves it untouched if the control is in the $\uparrow$ state.
Since the measurement ancilla is only influenced when the system qubit is in the $\downarrow$ state, by discarding results when the ancilla experiences a flip, we obtain the probability that the system was in the $\uparrow$ state.  By repeating the experiment with an anti-CNOT gate in which the role of $\uparrow$ and $\downarrow$ for the control qubit are switched, we obtain an \inm, and hence a non-invasive measurement of the state of the system.

\begin{figure}[tb]
  \begin{center}  
    \includegraphics[width=0.8\columnwidth,clip=true]{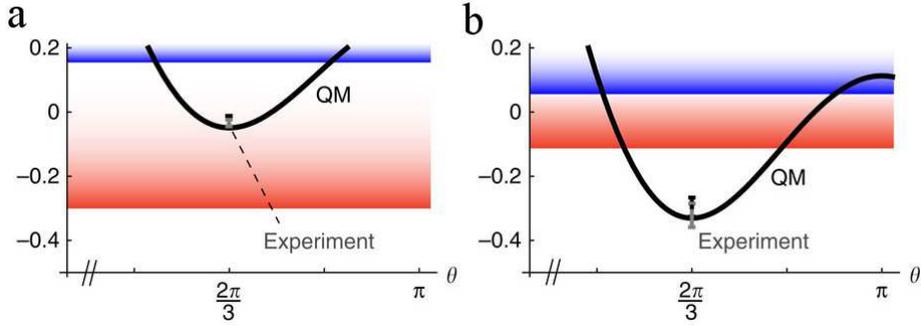}
  \end{center}
  \caption{
    (Color online) Results from the nuclear spin experiment of \citer{Knee2012}.
    Shown is the LGI correlator $f \equiv 1 - K_3'$ as a function of the system evolution angle $\theta$.  
    The black line shows the theoretical prediction and the data point, the experimentally-measured result.
    With perfect ancilla preparation, a realistic description of the spin implies that $f\ge 0$. Taking imperfect preparation into account, the bound on $f$ becomes \eq{fzeta} with  ``venality'' $\zeta$ the fraction of incorrectly-prepared ancillas, and this is shown in red (blue denotes a less strict bound, not discussed here). 
    The two figures show the results for two initial ensembles: (a) a thermal initial state (2.6~K) and (b) a highly polarized state. 
    For the thermal ensemble, the measured value lies between $-2\zeta$ and zero --- this means that the apparent violation of the LGI can be explained away in terms of ancilla errors. For the polarized ensemble, the measured value satisfies $f < -2\zeta$, such that a genuine violation of the LGI can be claimed.
    Figure from \citer{Knee2012}.
    \label{FIG:Knee_3AB}
  }
\end{figure}

To work as described, this measurement scheme requires that the ancilla be prepared in a pure state.
Without further consideration, deviations from exact purity could be exploited by a macrorealist to explain LGI violations.
To seal off this loophole, Knee \etal \cite{Knee2012} explicitly took the ancilla impurity into  account in their LGI tests. They considered the  quantity 
\beq
  f = 1 - K_3',
\eeq
which must be non-negative according to the standard Leggett-Garg arguments and under the assumption that perfect ancillas are used.  Knee \etal then define the ``venality'', $\zeta$, as the fraction of ancillas that are incorrectly prepared.  Taking into account incorrect preparation and assuming the worst case scenario,  they showed that their LGI must be modified to read 
\beq
  f\ge -2\zeta.
  \label{fzeta}
\eeq
The importance of this revised bound was demonstrated by considering two ancilla ensembles,  see \fig{FIG:Knee_3AB}.  Although results for a thermal ensemble at 2.6~K could violate the original bound, $f \ge 0$, the revised bound, \eq{fzeta}, was not violated --- the implication being that a macro-realist could plausibly ignore the conclusion of this experiment as an effect of unreliable measurement protocol.  However, by polarizing the ancilla such that the venality reached $\zeta=0.056$, even the revised bound could be violated (they measured a value of $f=-0.296$ as compared with a LGI lower bound of $-2\zeta = -0.112$) and thus a MR/NIM description could be ruled out.

With the precautions made in \citer{Knee2012} to ensure that their measurements were of the ideal negative type, as well as their mitigation of the ``venality-loophole'' which is, in principle, an issue for all measurement schemes using an ancilla, the work of Knee~\etal~\cite{Knee2012} represents the most complete experimental test of a LGI to date.

Katiyar \etal \cite{Katiyar2013} have also implemented an \inm scheme, similar to the foregoing, but with an NMR sample. LGI tests with liquid-state NMR systems rely on writing the state of the nuclear spin ensemble as
\beq
  \hat{\rho} = \epsilon \hat{\rho}_\mathrm{pure} + \frac{1}{2}(1-\epsilon) \hat{\mathds{1}}
  ,
\eeq
where $\rho_\mathrm{pure}$ is the pure part upon which the quantum operations are performed whilst the remaining maximally-mixed component remains unobserved in the experiment.  At room temperature, the parameter $\epsilon$ is very small --- \citer{Knee2012a} estimates a value of $\epsilon < 10^{-7}$ for the experiment of \citer{Souza2011}.  This gives rise to two problems. The first stems from the interpretation of the small $\epsilon$ as a low detector efficiency\cite{Knee2012a}.  To draw conclusions in the presence of such detector-efficiency requires the fair sampling hypothesis that the observed component is reflective of the entire ensemble to hold, which is not the case here.
Furthermore, it has been shown \cite{Menicucci2002} in the context of liquid-state NMR quantum computation that the results of quantum operations on a small number of liquid-state NMR spins can always be described in terms of a local hidden-variables theory.  Menicucci and Caves conclude that ``...NMR experiments up to about 12 qubits cannot violate any Bell inequality, temporal or otherwise''.
Souza~\etal seem to agree and write in \citer{Souza2012} that ``... [their] experiment can only be viewed as a demonstration of the circuit and not as a disproof of macroscopic realism'' and go on to state that the same conclusion should apply to other NMR experiments such as \citer{Athalye2011}, and by extension, \citer{Katiyar2013}.  Thus, whilst Katiyar \etal \cite{Katiyar2013} rightly seek to close off the clumsiness loophole with their use of \inm, the loopholes intrinsic to NMR prevent a serious challenge to a macrorealistic description of nature.


The experiment by George~\etal~\cite{George2013} draws the connexion between the violation of a LGI and winning quantum strategies in a quantum game.  This in itself is an interesting point, but the experiment is also important as it represents the only LGI test to date where the system is of greater complexity than a single qubit.  
Unfortunately, some of the discussion accompanying their results adds unnecessary confusion (much of which has been addressed by one of the authors \cite{Maroney2012}).

The quantum game in question is the three-box game \cite{Aharonov1991}, played by two protagonists, Alice and Bob, who manipulate the same three-level system. 
We will just describe the quantum sequence of events for this game, and refer the interested reader to the above articles for the details and classical play of this game.
Alice first prepares the system in state $\ket{3}$, and then evolves it with a unitary operator that takes 
\beq
  \ket{3} \to \frac{1}{\sqrt{3}}\rb{\ket{1} +\ket{2}+\ket{3} }
  \label{UI}
  .
\eeq
Bob then has a choice of measurement: with probability $p^B_1$ he decides to test whether the system is in state $\ket{1}$ or not (classically, he opens box 1), and with probability $p^B_2$ he tests whether the system is in state $\ket{2}$ or not.  Alice then applies a second unitary to the system, which takes
\beq
  \frac{1}{\sqrt{3}}\rb{\ket{1} +\ket{2}-\ket{3} } 
  \to 
  \ket{3}
  \label{UF}
  ,
\eeq
before she makes her final measurement to check the occupation of state $\ket{3}$.

If both Alice and Bob find the system in the state that they check (e.g., Bob measures level 1 and finds the system there and Alice, the same for state 3), then Alice wins. 
If Alice finds the system in state 3, but Bob's measurement fails, then Bob wins. Finally, if Alice doesn't find the system in state 3, the game is drawn.
In a realistic description of this game in which Bob's measurements are non-invasive, Alice's chance of winning can be no better than 50/50 as long as Bob chooses his measurements at random ($p^B_1 = p^B_2=1/2$).  
In the quantum version as described above, however, interference between various paths means that Alice wins every time.  Alice's quantum stategy therefore outstrips all classical (i.e. realistic, NIM) ones.

George~\etal realised this three-box quantum game in a nuclear spin system. They then went on to consider a LGI, $K_3'$ of \eq{K3'}, for the system, where Alice's preparation constitutes the first measurement, Bob's measurement the second, and Alice's final measurement, the third.
In  evaluating the LGI, all measurements assign a value of $+1$ to state 3 and a value $-1$ to the other states. 
The direct way to implement such a measurement would simply be to use the two projectors $\op{3}{3}$ (value $+1$) and $\rb{\op{1}{1} + \op{2}{2}}$ (value $+1$).  
Using this projective measurement scheme and unitary evolutions consistent with \eq{UI} and \eq{UF}, 
the three correlation functions obtained in the two-point fashion evaluate as $\ew{Q_2 Q_1} = -1/3$, $\ew{Q_3 Q_2}=-3/9$ and $\ew{Q_3 Q_1}=-7/9$.  The LG parameter $K_3'$ in turn evaluates as
\beq
  K_3' = \frac{13}{9} > 1
  \label{three-boxviol}
  ,
\eeq
which represents a clear violation of the LGI.

This measurement scheme, however, is not the one pursued by George~\etal.
%
Rather, they restrict the middle measurements to those permitted to Bob in the three-box game: a 
yes/no measurement for whether the system in state 1, and yes/no measurement for whether the system in state 2.  
Under a realistic understanding of the system, knowledge of the probabilities of the outcomes of these measurements allow one to construct the probability that the system was in state 3.  
George~\etal are therefore able to obtain $K_3'$ using only that set of measurements involved in the three-box game.  The calculated value of $K_3'$ using these measurements is exactly as in \eq{three-boxviol}.  In the experiment, a value of $K_3' = 1.265 \pm 0.23$ was measured which, although  differing from the theoretical expectation, still shows a clear violation of the inequality.
The significance of this result is that the authors were able to show that violation of the LGI necessarily implies that the corresponding quantum strategy adopted by Alice will allow her to win the three-box game with a probability higher than classical theory will allow.  This result therefore suggests a general correspondence between LGI violations and winning quantum strategies.

In discussing their results, the authors of \cite{George2013} introduce the concept of ``non-disturbing measurements''.  As formulated in \cite{Maroney2012}, a measurement of $Q_2$ at time $t_2$ is non-disturbing (from the point of view of a subsequent measurement at time $t_3$) if
\beq
  P_3(Q_3) = \sum_{Q_2} P_{32}(Q_3,Q_2)
  ,
\eeq
i.e., the results of the measurement at time $t_3$ are the same irrespective of whether the measurement at $t_2$ is performed or not (we will meet this concept again in \secref{SEC:related} under the guise of a quantum witness or no-signalling in time).
In the three-box game, it can be shown that Bob's measurements do not disturb Alice's later results.
Unfortunately, in \citer{George2013}, the concept of a non-disturbing measurement is conflated with that of a non-invasive one, although the distinction between the two 
is made in \citer{Maroney2012}.
In the language of this review, and indeed most of the literature on the topic, the ``non-disturbance'' character of the measurement is equivalent to saying that the measurement is a weak measurement.
That this is so can be seen by observation that, if Bob's measurements are non-disturbing, then in the LGI it does not matter whether the correlation functions are measured in the same, or in separate, runs. Thus, two-point and three-point ways of obtaining the LGI are equivalent, which is only the case if the measurements are weak (see \secref{SEC:2point}). 
Indeed, we would categorise the LGI test of the three-box protocol as a three-point LGI test where the middle measure is weak, consisting of a POVM defined by the set of projectors $p^B_1\op{1}{1}$, $p^B_1\rb{\op{2}{2} + \op{3}{3}}$, $p^B_2\op{2}{2}$, and $p^B_2\rb{\op{1}{1} + \op{3}{3}}$. 
A consequence of this interpretation, when combined with the connection between three-point LGI violations and the existence of weak values \cite{Jordan2006,Dressel2011}, is that better-than-classical quantum strategies should also be associated with weak values. This is indeed found to be the case in e.g.~\citer{Ravon2007}.

Finally, we note that proof is given in \citer{George2013} that, for two-level systems, violation of a LGI necessarily means that the measurements are disturbing.  
As can be seen from the numerous examples of LGI violations for two-level systems measured in a three-point weak-measurement (non-disturbing in this language) fashion, this is not true in general, but holds only if the measurements are assumed to be  projective measurements acting directly on the system itself. 
Furthermore, if we understand this result to apply for projective measurements on the system, the result is trivially extended to arbitrary system size 
--- since non-disturbing (weak) measurements imply the equivalence of two-point and three-point LGIs, and we know that three-point LGIs with projective measurements can not yield violations (see section \secref{SEC:3point}), then projective measurements that give a LGI violation must be disturbing.
What is interesting about the three-box problem is that the partial projections performed on the system by Bob in a probabilistic fashion enable him to build a POVM that implements a weak measurement on the system. Since partial projections require more than two levels, this only becomes possible once the system has a Hilbert space larger than that of a qubit. It is interesting to note that e.g. Goggin~\etal~\cite{Goggin2011} enact their weak measurement by adding an auxiliary qubit to the system and making projective measurements in this extended Hilbert space.

\section{Light-matter interactions\label{SEC:crystal}}

Aside from superconducting qubits and nuclear spins, the only other report of a violation of a LGI in a matter system is the work of Zhou \etal\cite{Zhou2012}.  Their system consisted of
two millimeter-scale pieces of Nd$^{3+}$:YVO$_4$ crystal separated by a half-wave plate.
Using an atomic-frequency comb technique, they could tailor the absorption spectrum of the crystal so that a single input photon created a state in one of the crystals of the form
\beq
  \ket{e}_N = \sum_j^N c_j e^{-i k z_j} e^{i2\pi \delta_j t} \ket{g_1\centerdot\centerdot\centerdot e_j \centerdot\centerdot\centerdot g_N}
  ,
\eeq
where $N \sim 10^3$ is the number of atoms involved in the delocalized excitation; $g_j$ ($e_j$) indicates that atom $j$ (with position $z_j$) is in the ground (excited) state;  $k$ is the wavenumber of the input field; $\delta_j$ is the detuning between atom and input laser frequency; and $c_j$ is an atom-dependent amplitude.
This state, similar to a Dicke- \cite{Dicke1954} or $W$-state \cite{Dur2000}, also arises in arrays of quantum wells, and has been discussed in terms of a LGI violation in Chen \etal~\cite{GYSciRep}.

By simultaneously illuminating both crystals and tuning the phase $\psi_0$ of the polarization $H+V\exp\rb{i\psi_0}$ of the input photon, the crystals can be prepared in the joint state,
\beq
 \psi(t) =
 \frac{1}{\sqrt{2}}
 \left\{\ket{e}_{N1}\ket{g}_{N2} + \ket{g}_{N1}\ket{e}_{N2}\exp{\left[i(2 \pi \delta t + \psi_0)\right]} \right\},
\eeq
where $\delta$ is the frequency detuning between the two atomic frequency combs.
The Leggett-Garg measurement was chosen as a measurement in the basis 
  \beq \ket{D}=\frac{1}{\sqrt{2}}\left(\ket{e}_{N1}\ket{g}_{N2} + \ket{g}_{N1}\ket{e}_{N2}\right)
  ,
\eeq
with eigenvalue $+1$ and 
  \beq \ket{A}=\frac{1}{\sqrt{2}}\left(\ket{e}_{N1}\ket{g}_{N2} - \ket{g}_{N1}\ket{e}_{N2}\right)
  ,
\eeq  
with eigenvalue $-1$.  The occupation of states $\ket{D}$ and $\ket{A}$ was measured as a function of time by the observation of the polarization state of an emitted photon at some time after the state was created.  This measurement set-up, which involves state-preparation followed by an invasive-measurement meant that Zhou \etal~\cite{Zhou2012} investigated Huelga's inequality of \eq{LGI_Huelga}. Thus, the observed violations can, at best, be associated with the lack of a Markovian description of the system.  As discussed by Chen \etal \cite{GYSciRep} performing a test of the standard LGIs on a Dicke- or W-state is challenging

The issue of whether this system exhibits macroscopic coherence or not is an interesting one.
While the crystals are certainly macroscopic, both in size and separation, and the excitation consists of a coherent distribution of phase between a macroscopic number of particles, at the end of the day, the interfering states only differ by the absorption of a single quantum.  Correspondingly, the disconnectivity, $\mathfrak{D}$, and extensive difference $\mathfrak{L}$ are small.  The situation is thus similar to the experiment of Palacios-Laloy, in that we should talk here of a test of microscopic coherence in a macroscopic system.

We also note that Sun \etal \cite{Yong-Nan2012} have proposed a test of \eq{LGI_Huelga} via an optical excitation of biexciton states in a single quantum dot.

\section{Optics\label{SEC:optics}}
A single photon is perhaps as far from being a macroscopic object as one can imagine.  Nevertheless, tests of the LGI with photons have attracted significant interest, particularly in connexion with weak measurements.

The simplest optical LGI test would be the Mach-Zehnder interferometer \cite{Kofler2013,Emary2012}, in which the arm index is taken as the system's qubit degree of freedom and the time-evolution of the particle is generated by beamsplitters. Measuring $K_3$ requires two beamsplitters with the measurement times $t_i$ mapped onto positions in the interferometer: $t_1$-measurements are made before the first beamsplitter, $t_2$-measurements between them, and those for $t_3$ are made at the output ports.  
Measurements at these points can be made by inserting  photo-detectors into the arms of the interferometer and this presents a very natural way to realise an \inm \cite{Emary2012}, as a macroweaselist would have to claim that the photon taking one path was affected by the presence of a detector in the other.

Another simple optics set-up is that considered experimentally by Xu \etal\cite{Xu2011}, where the observable $Q$ was the polarisation of a single photon.  The set-up was similar to that of \fig{FIG:Knee} with a single photon ancilla and CNOT gate to perform the middle measurement (no account of non-invasiveness was given, though).  The time-evolution ($U_{ij}$ in \fig{FIG:Knee}) was produced by angled quartz plates that induced a relative phase between polarisation components.  The frequency dependence of this phase combined with the spread of the initial wave packet was used to simulate the dephasing effects of an environment.
As noted in \citer{Xu2011} and \citer{Emary2012}, a classical laser pulse would violate the LGI in both this and the Mach-Zehnder set-up, since classical wave mechanics is not a macroscopic-real theory.  This is a reminder that the violations of a LGI cannot strictly-speaking be taken as evidence of quantum mechanics, but rather evidence of the absence of a description along the lines of \A{1-3}.

\subsection{Optical LGIs and weak measurements \label{SEC:optics_weak}}
\begin{figure}[tb]
  \begin{center}  
    \includegraphics[width=0.9\columnwidth,clip=true]{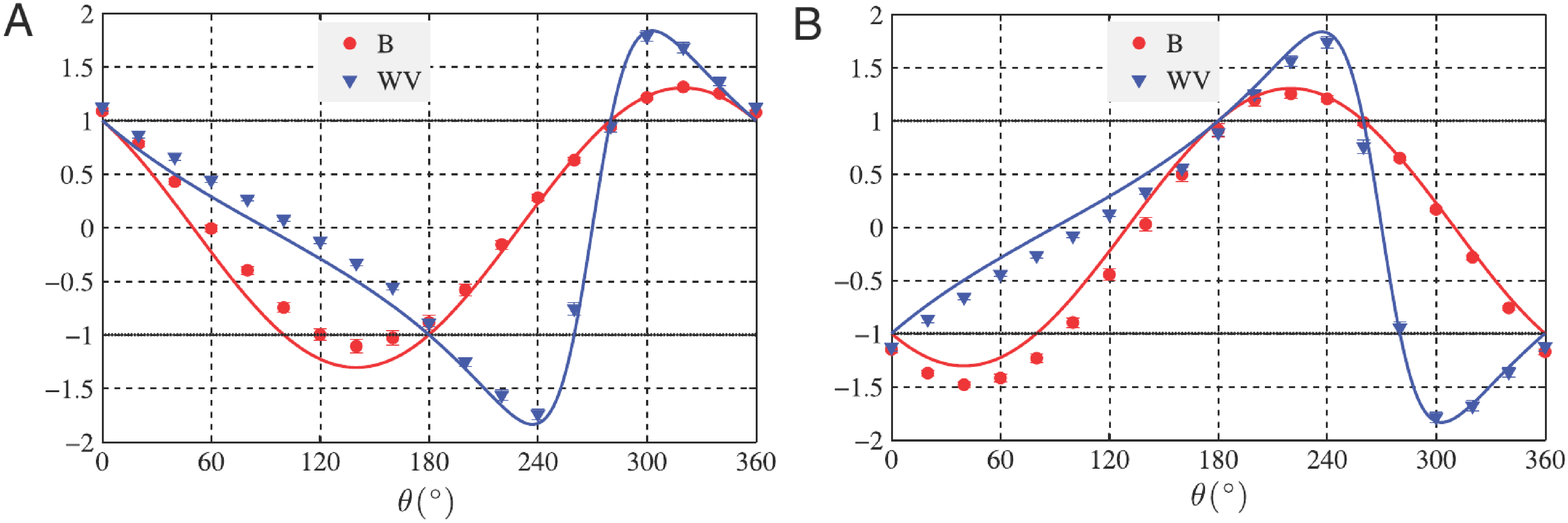}
  \end{center}
  \caption{
    (Color online) Results from the optics experiment of \citer{Goggin2011}, which show the Leggett-Garg parameter (labelled B, red) and a weak value (labelled WV, blue) for a range of input states parametrized by the angle $\theta$.  
    The two parts show results for different measurements at the second position in the three-measurement experiment, with 
    solid lines showing the theoretical predictions and points, the experimental data.
    It is evident that for this 3-point LGI test with semi-weak measurements, a violation of the LGI is accompanied by the emergence of a weak value.  The measurement strength here was $K = 0.5445 \pm 0.0083$.  The experiment was repeated with $K = 0.1598 \pm 0.0091$ and larger LGI violations were observed.
    Figure from Goggin~\etal~\cite{Goggin2011}.
    \label{FIG:Goggin}
  }
\end{figure}

LGI tests with weak measurements have been performed in several optical set-ups \cite{Goggin2011,Dressel2011,Suzuki2012}.  Goggin \etal \cite{Goggin2011} considered a polarisation qubit in a system-ancilla configuration somewhat similar to \fig{FIG:Knee}.  The state of the system qubit at $t_1$ is simply defined as the $Q=+1$ state; the operator $U_{10}$ was absent, and $U_{21}$ was chosen such that its output state was
\beq
  \ket{\sigma} =  \cos \theta/2 \ket{H} + \sin \theta/2 \ket{V},
\eeq
with $\ket{H,V}$ two orthogonal linear-polarisation directions; $U_{32}$ was chosen such that the measurement at $t_3$ is effectively measured in the 
basis
\beq
  \ket{D,A} = 2^{-1/2}\rb{\ket{H} \pm \ket{V}}
  .
\eeq
The inequality that was measured was therefore
\beq
  K_3 = \ew{Q_2} + \ew{Q_2 Q_3} -\ew{Q_3} \le 1
  ,
\eeq
with operators $\hat{Q}_2 =\hat{\sigma}_z$ and $\hat{Q}_3 = \hat{\sigma}_x$  (see \secref{SEC:EiT}).
The measurement at $t_2$ was performed with a C-SIGN gate in which the $\ket{VV}$-component of the system-ancilla wavefunction obtains a phase-inversion.
The ancilla photon was prepared in the pure superposition state $\rho_A = \op{\mu_\mathrm{in}}{\mu_\mathrm{in}}$ with 
\beq
  \ket{\mu_\mathrm{in}} = \gamma \ket{D} + \bar{\gamma}\ket{A},
\eeq
and $\gamma^2 + \bar{\gamma}^2=1$.  This superposition of ancilla states allows one to alter the type of measurement made at $t_2$: for $\gamma = 1$, the measurement is strong and performs an ideal projective measurement of the system; for $\gamma \to 1/\sqrt{2}$, the measurement is weak with minimal information gathered per run. Goggin \etal\cite{Goggin2011} describe this range of possibilities by the parameter (``knowledge'') 
\beq
  K \equiv 2\gamma^2 -1
  ,
\eeq
ranging from $K=1$ for a strong measurement and $K=0$ for a weak one.
The measurement at $t_3$ is performed projectively. Both $Q_2 $ and $Q_3 $ were measured in every run (three-point measurement) and the probabilities of detecting system and ancilla photons in their various states were obtained.
Based on the knowledge of the detector action, Goggin \etal determined the expectation value of $Q_2$, as obtained by the weak measurement, as
\beq
   \ew{Q_2} = \frac{P_a(D)-P_a(A)}{K}
   ,
\eeq
where $P_a(X=D,A)$ is the probability to find the ancilla in state $A$ or $D$ at the final measurement. 
Based on this, they reported violations of the LGI for two different values of the measurement parameter, $K$, with a larger violation associated with the smaller $K$-value (weaker measurements). Results from this experiment are shown in \fig{FIG:Goggin}.

Suzuki~\etal~\cite{Suzuki2012} also considered a polarisation qubit with three-point measurements, but they implemented the measurement of $Q_2$ with an interferometer set-up \cite{Iinuma2011}, which allows a complete tuning from weak to strong measurements. With the qubit initialised in the $Q=+1$ state they showed that the  probabilities $P_\mathrm{exp}(Q_2,Q_3)$ obtained directly in the experiment do not violate a LGI. 
From \eq{K3probs}, we see that this would require $P_\mathrm{exp}(-1,+1)$ to be negative. 
Indeed, Suzuki~\etal~interpret the lack of LGI violations with the raw measured probabilities as being ``because the errors in measurement resolution and back-action required by the uncertainty principle guarantee that $P_\mathrm{exp}(-1, +1)$ will always remain positive''.
This statement strongly echoes the arguments made by Onofrio and Calarco \cite{Onofrio1995,Calarco1995,Calarco1997,Calarco1999}, who have consistently argued against the observability of LGI violations due to the uncertainty principle. 
Whilst Onofrio and Calarco maintain that their argument applies to the original, two-point method of measuring the LGI tests \cite{Onofrio2013}, we find that it only makes sense when restricted to the three-point method with projective measurements, in line with Suzuki \etal in the above quote.

Suzuki~\etal~went on, however, and by introducing a model for their detector which takes into account the finite resolution of a weak measurement, they obtained revised quasi-probabilities  such that, as the measurement became weaker, the relevant quasi-probability $P(-1, +1)$ became negative and thus LGI violations were observed.
Suzuki \etal \cite{Suzuki2012} also included a classical back-action effect in their detector model. By including the two effects (finite resolution and this back-action) they arrived at a quasi-probability $P(-1, +1)$ that was both negative and independent of measurement strength.  They concluded therefore that this negative probability is inherent to the original state and not dependent on the type of measurement performed.  
While this may be the case, since Suzuki \etal \cite{Suzuki2012} consider that their detectors are producing a classical back-action effect, no conclusions regarding the LGI can be made, due to the conflict with the NIM assumption.


\begin{figure}[tb]
  \begin{center}  
    \includegraphics[width=0.5\columnwidth,clip=true]{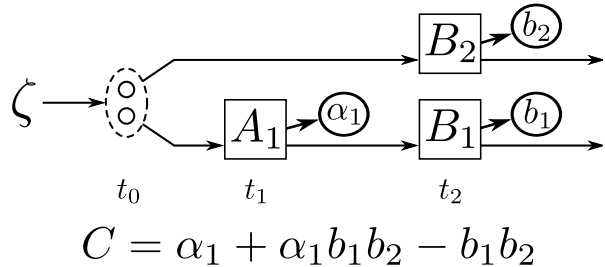}
  \end{center}
  \caption{
    A two-party Leggett-Garg inequality with measurements as discussed in \citer{Dressel2011}.
    A pair of particles is extracted from the ensemble $\zeta$ and then subjected to the measurements $A_1$, $B_1$ and $B_2$ in the sequence shown, yielding measurement results $\alpha_1$, $b_1$, $b_2$.  Measurements $B_i$ are projective, whereas measurement $A_1$ is semi-weak.
    A Leggett-Garg inequality is then investigated for the two-party correlator $C \le 1$.
    Figure from \citer{Dressel2011}.
    \label{FIG:Dressel}
  }
\end{figure}

Dressel~\etal~\cite{Dressel2011} derived and tested a novel weakly-measured LGI in which the system under test was a pair of particles. 
The sequence of measurements on the particle pair is illustrated in \fig{FIG:Dressel} in which detector $A_1$ may be ambiguous (corresponding to a semi-weak measurement in the quantum case) and the end detectors $B_1$ and $B_2$ are unambiguous (corresponding to projective measurements in the quantum case).  In each run, the three detectors obtain values $\alpha_1$, $b_1$, and $b_2$ and the quantity 
\beq
  C=\ew{A_1 + A_1 B_1 B_2 - B_1 B_2}
  ,
\eeq
constructed.
With a derivation similar to that given in \secref{SEC:3point}, Dressel~\etal  showed that this quantity is bounded $-3 \le C \le 1$ under MR and under the assumption that the detector $A_1$ is both non-invasive and unambiguous.

The set-up in \fig{FIG:Dressel} was implemented with polarization qubits with  measurement $A_1$, a semi-weak one.  Violations were observed in line with quantum theory.  
An interesting aspect of the experiment is that in order to obtain violations, the two particles had to be entangled with one another. This suggests that the inequalities of Dressel \etal combine aspects of both Bell and Leggett-Garg inequalities. Another work that appears to span these two types of inequality is \citer{Milman2007}.

A focus of the experiments in \citer{Goggin2011} and  \citer{Dressel2011} was to investigate the prediction of Williams and Jordan \cite{Williams2008,Williams2009} that the violation of every ``generalised LGI'' (i.e. one measured with weak measurements) can be associated with the occurrence of a so-called {\em strange weak value} \cite{Aharonov1988,Kofman2012} for a system variable.  Strange weak values are measured values of an observable that lie outside the eigenspectrum of the observable.  They can arise under the conditions of weak measurement and post-selection and have a ``...long history of controversy...'' \cite{Dressel2010}, which we shall not go into here (see citations in Refs.~\cite{Aharonov2001,Dressel2010}).
FIor example, in conjunction with their LGI experiment, Goggin \etal looked at weak values such as
\beq
    \mathstrut_D\ew{Q_2} =\frac{P_{a|s}(D|D)-P_{a|s}(A|D)}{K}
    ,
\eeq
where, e.g., $P_{a|s}(A|D)$ is the conditional probability of finding the ancilla photon in state $\ket{A}$ given that the system is found in state $\ket{D}$.
Considering a range of weak values and the (third-order) LGIs for this problem, they indeed find a strange weak value whenever a generalised LGI is violated, see \fig{FIG:Goggin}.
The violations of the two-particle LGIs \citer{Dressel2011} were likewise associated with strange weak values and their conjunction understood in terms of contextual values \cite{Dressel2010,Dressel2012}.

\section{Quantum transport \label{SEC:QT}}

Quantum transport studies the motion of electrons through structures small enough in dimension that the quantum nature of the electron plays an important role \cite{Brandes, Kouwenhoven}.
Such systems show a rich interplay between non-equilibrium and quantum physics, which has been revealed through both time-resolved charge and more standard transport measurements, such as current and noise (e.g.,~\cite{ Kouwenhoven, Shinkai, Kiesslich, Petta05sci,PNAS, FujiQPC}).
To date, there have been several theoretical works investigating the possibility of observing  violations of LGIs in transport systems.

In \citer{Lambert2010}, Lambert \etal~first considered how a LGI can be violated by measuring the location of the electron charge in some discrete region within a quantum nanostructure.
The assumption of Coulomb blockade provides an upper bound for the charge in the system, such that one can define a bounded operator as required by the LGIs.
Let us assume a charge detector that registers the value $Q'_n\geq 0$ when the system is in the $n$th of $N$ possible states and that state $N$ is the state for which  $Q'$ has its maximum value: $Q'_N = Q'_\mathrm{max}$.
Then, defining the bounded operator $Q= 2Q'/Q'_\mathrm{max}-1$ and introducing this into the stationary three-term LGI of \eq{LGnterm_stat}, one obtains,
\beq
  2\ew{Q'(t)Q'} - \ew{Q'(2t)Q'} \leq Q'_\mathrm{max} \ew{Q'}
  \label{LGcharge}
  .
\eeq
The use of the stationary LGI here is motivated by the fact that in transport experiments this is typically the regime of interest. Lambert \etal \cite{Lambert2010} also showed that this inequality can hold in the non-stationary regime (i.e., with arbitrary initial states) but only under the conditions of a Markovian, time-translationally-invariant evolution and when only a single state contributes to the detection process, i.e., $Q_n =
Q_\mathrm{max} \delta_{nN}$.  In this second case, the equality is then similar to that of \eq{LGI_Huelga}.
Lambert \etal went on to show theoretically the violation of \eq{LGcharge} (in the stationary case) by measurements of the position of a single electron within a double quantum dot in the large bias, Coulomb Blockade, regime.  The effects of a phonon bath were included, and even though this damped  the oscillations of the LGI correlator,  violations at short times were found to remain up to relatively large phonon temperatures.
Lambert \etal also derived an additional inequality for the current flowing through the double quantum dot. Although the instantaneous current is an unbounded observable and a simple LGI of the form \eq{LGcharge} cannot generally be constructed, under some additional, rather strict, assumptions pertinent to the double quantum dot in the large bias regime, just such an inequality was derived and shown to be violated by the quantum description of the problem. This same inequality has also been discussed in terms of photonic `current' measurements in cavity-QED systems \cite{Lambert10PRA}.

A direct measurement of \eq{LGcharge} would prove difficult in practice due to the short time-scales over which the correlation functions need to be measured (of the order of a nanosecond \cite{Hayashi2003}). Moreover, it may be difficult to construct charge measurements that satisfy the NIM criterion.
Emary \etal \cite{Emary2012a} proposed electron interferometers as a way to overcome these  difficulties.  The simplest set-up they considered was an electronic Mach-Zehnder interferometer realised by quantum Hall edge-channels.  The test of the LGI, \eq{K3intro}, in this system proceeds in direct analogy with the photonic Mach-Zehnder interferometer discussed in \secref{SEC:optics}, with single electrons in edge channels replacing photons propagating in free space, and quantum point contacts playing the role of beamsplitters.  The advantages of this Mach-Zehnder geometry is that it enables the unambiguous implementation of \inm and only mean currents, rather than time-dependent correlation functions, need to be measured.

\subsection{Full counting statistics}
Full counting statistics seeks to understand electronic transport by counting the number of charges transferred through a conductor in a certain time interval $t_b\ge t\ge t_a$ \cite{Levitov1996,Levitov2002}.
Considered as a {\em classical} stochastic process, the information about transferred charge can be encapsulated by the moment-generating function
\beq
  \mathcal{G}_\mathrm{cl.}(\chi;t_b,t_a) = \ew{\exp{\left[i\chi( n(t_b) - n(t_a) )\right]}}
  \label{MGFcl}
  ,
\eeq
where $n(t)$ is the collector charge at time $t$ and $\chi$ is the counting field.
In  \citer{Emary2012a}, Emary \etal showed that the quantity
\beq
  L(\chi,\left\{t_i\right\})
  \equiv
  \mathcal{G}(\chi;t_1,t_0)
  +
  \mathcal{G}(\chi;t_2,t_1)
  -
  \mathcal{G}(\chi;t_2,t_0)
  \label{Ldefn}
  ,
\eeq
which involves the moment-generating function over three different time intervals obeys the Leggett-Garg-inspired inequalities
\beq
  B_\mathrm{R}(\chi)
  \le
  \mathrm{Re}\!
  \left\{
    L(\chi,\left\{t_i\right\})
  \right\}
  \le C_\mathrm{R}(\chi)
  ;
  \\
  - C_\mathrm{I}(\chi)
  \le
  \mathrm{Im}\!\!
  \left\{
    L(\chi,\left\{t_i\right\})
  \right\}
  \le C_\mathrm{I}(\chi)
  ,
  \label{FCSineq}
\eeq
for all $\chi$ and times $\left\{t_i\right\}$. These inequalities were derived under the usual Leggett-Garg assumptions, \A{1-3}, with the additional assumption of charge quantisation.  In these inequalities, the bounds are $\chi$-dependent with, for example, $C_\mathrm{R}(\chi) = 1$, $B_\mathrm{R}(\chi) = -3$ and $C_\mathrm{I}(\chi) = 0$ when $\chi=\pi$, corresponding to a parity measurement of the reservoir charge.

The canonical quantum-mechanical moment-generating function of full counting statistics was given by Levitov and coworkers \cite{Levitov1996, Levitov2002} as
\beq
  \mathcal{G}_\mathrm{L}(\chi;t_b,t_a)
  &=&
  \ew{
    \exp{\left[-i\frac{\chi}{2} \hat{n}(t_a)\right]}
    \exp{\left[\frac{}{}i\chi\hat{n}(t_b)\right]}
    \exp{\left[-i\frac{\chi}{2} \hat{n}(t_a)\right]}
  }
  \label{MGFlevitov}
  .
\eeq
The set-up proposed to measure this moment-generating function was a spin processing under the influence of the magnetic field generated by the collector current.  In this set-up, the counting field $\chi$ has the physical significance of being the coupling strength between system and detector and so can, in principle, be scanned through. In the ideal case, this measurement can be performed non-invasively.
In \citer{Emary2012a} it was shown that both normal-metal and superconducting  single-electron transistors can cause violations of inequalities \eq{FCSineq}.

The inequalities \eq{LGcharge} and \eq{FCSineq} are complimentary to one another ---
the former tests the existence of a macrorealist description of charges inside a nanostructure,  the latter tests the same for the charges in the leads. Correspondingly, the former can be violated by quantum superpositions within the structure, whereas the latter can be violated by coherences between the system and the lead.
Whilst these inequalities were derived for charge flow in quantum transport, this approach should be applicable to any dynamical stochastic process.
The significance of inequalities of the form \eq{FCSineq} is that they give classical bounds based on the complete statistical information about the system, which can be arbitrarily complex (i.e. they are not just restricted to $Q=\pm 1$ observables).  In this sense they are similar to the entropic LGIs.

Bednorz and Belzig  \cite{Bednorz2010a} have considered theoretically continuous weak measurement of the current through a mesoscopic junction and derived an inequality, similar in spirit to the LGIs, but involving up to fourth-order current cumulants in the frequency domain  \cite{Bednorz2010}. Violations of this inequality were obtained for a quantum point contact. A related fourth-order inequality was discussed for a qubit in \cite{Bednorz2012}.

\section{Photosynthesis\label{SEC:photosynth}}


The possible role of quantum coherence in certain
biological functions has garnered a great deal of interest in the
last decade. In 2007 Engel \etal~\cite{EngelNature07,PNAS}
performed an experiment, on a particular pigment-protein complex from
the light-harvesting apparatus of green sulphur bacteria, which
revealed the apparent wave-like quantum coherent motion of a
single electronic excitation through the complex. This complex, termed the
Fenna-Matthews-Olson (FMO) complex, consists of seven
``bacteriochlorophyll a" molecules, which in totality act as a wire
connecting a large antenna complex to the reaction center. Photons
are absorbed by the antenna complex as electronic excitations, and
are then routed through an FMO trimer to the reaction center. The
highly efficient transfer of these excitations has been the
subject of much discussion, and the possible role of quantum
coherence in enhancing this efficiency has played a fundamental
part in the development of the field of quantum biology
\cite{Lambert12}.

The observation of coherent oscillations
\cite{EngelNature07,PNAS} is intriguing. However, it
has been often argued that a variety of other phenomena could
induce similar signatures. To help resolve this
argument Wilde \etal \cite{Wilde2010} proposed the
application of an LGI to the FMO complex, in
the spirit of using an LGI as a  tool to verify the presence of
quantum coherence and eliminate other ``classical'' explanations
of the wave-like phenomena.  In their work they calculated $K_3$ and its cousins and found the timescales on which a violation
may be observed under certain assumptions about the environment.

A practical implementation of such a phenomenon seems
difficult at this time. Experiments on the FMO complex so far rely
on two-dimensional spectroscopy, which does not
correspond to an idealised measurement in the site basis, and is presumably highly invasive. In addition, even at $77$ Kelvin the violation of the LGI occurs only on a timescale of $0.035$ ps \cite{Cheming} (the value in \citer{Wilde2010} differs), which may be exceptionally difficult to observe. Li \etal \cite{Cheming}
(and independently Kofler and Brukner \cite{Kofler2008,Kofler2013}) proposed
an alternative to the LGI (see \eq{Qwit} in section \secref{SEC:related} for a full discussion) which gives a
broader window of violation ($t_0=0.3$ ps at $77$ K, based on a model
of the FMO complex which included strong
coupling to a non-Markovian environment).  However, an unambiguous test of the quantum
coherence, with an LGI or otherwise, remains to be realized experimentally.

\section{Nano-mechanical systems\label{SEC:NEMS}}
Nano-mechanical systems are mechanical oscillators fabricated on
the nano-scale~\cite{BlenPR,Poot2012}. Such devices come in several varieties, including
single- and doubly-clamped semi-conductor beams, cantilevers, toroidal, and drum geometries.
They are typically characterized by an exceedingly
high frequency of oscillation $\omega_m$ (of the order of
giga-Hertz) and large quality factor $Q$. In several
experiments~\cite{oconnell,teufel} such devices have been cooled to
temperatures low enough ($k_B T \ll \hbar \omega_m$) that
the quantum ground state motion of their
centre of mass can be observed, and potentially manipulated.

As pointed out by several authors\cite{Wei2006,clerk10, clerk11} there remains an
ambiguity in distinguishing whether the ground state motion of such systems is quantum or classical,
particularly in opto-mechanical setups like Teufel {\em et
al.}~\cite{teufel}. This ambiguity arises because both quantum mechanics and classical mechanics
predict nearly identical properties for linear harmonic oscillators. The only easily accessible quantum signature in this case
is the non-zero quantum vacuum displacement of a harmonic oscillator as $T\rightarrow 0$ (though the asymmetry in the spectral properties
of absorption and emission of quanta has been identified as a purely quantum effect and observed in experiments~\cite{Painter2012}).
Can the Leggett-Garg inequality assist in this
case? Naive considerations say no, as the measurement of both the
displacement and the energy of a nano-mechanical system are
unbound continuous  variables, which do not satisfy the Leggett-Garg requirement of
bound or dichotomic observables. However, two approaches have been
suggested. The first is to construct a dichotomic measurement
using dispersive coupling to a qubit~\cite{LambertNem, Johnson}.
This dispersive coupling allows one to distinguish between the
occupancy of different vibronic states within the mechanical
system, allowing one to perform the effective dichotomic
measurement 
\beq 
  \hat{Q}_m = 2\ket{1}\bra{1} -\mathds{1}
  .
\eeq  
In other words, one could
measure whether there is one phonon in the mechanical system, or
not. The second approach is to consider the extended class of
inequalities for continuous variable measurements such as in \citer{Bednorz2011}.  An initial examination of this
possibility was also discussed by Clerk~\cite{clerk11}, and Lambert \etal~\cite{LambertNem}.
However, no explicit proposal showing how such higher-order correlation functions could be measured on
a nano-mechanical device has been made, and even conceptualizing such an implementation remains challenging.

\section{Related tests of macrorealism \label{SEC:related}}
Whilst we have restricted the scope of this review to the LGIs, or their very close relatives, there exist a number of related tests of macrorealism that are worth comment.

In analogy with Bell's theorem without inequalities \cite{Greenberger1990,Hardy1993,Mermin1993}, a number of authors have written down equalities based on the Leggett-Garg assumptions \cite{Foster1991,Jaeger1996}, although at least some of these appear to be unmeasureable \cite{Elby1992}.
%
In two recent works Li \etal \cite{Cheming}
and, independently Kofler and Brukner~\cite{Kofler2013} (in the spirit of their earlier discussion \cite{Kofler2008}) proposed
an alternative to the Leggett-Garg inequality based on the same macrorealism assumptions of the LGI.  Assumption \A{1} implies that, since the system will have a well-defined value of $Q$ at times where it is not measured, the probabilities used to determine measurement results can be obtained as the marginal of a two-time probability distribution (which is itself a marginal of a three-time probability distribution),
\beq
  P_{i}(Q_i) = \sum_{Q_k} P_{ik}(Q_i,Q_k)
  \label{Qwit}
  ,
\eeq
which was called the ``no-signalling in time'' condition in Refs.~ \cite{Kofler2008,Kofler2013} and a ``quantum witness'', in analogy to entanglement witness \cite{Horodecki09}, in \citer{Cheming}. 
The main result of these two works is to suggest that deviations from this equality can be used as a test of macrorealism plus NIM directly.
This criterion was also described as the ``non-disturbing measurement'' criterion in \citer{George2013}.

Both Li~\etal~\cite{Cheming} and Kofler and Brukner~\cite{Kofler2013} showed that this witness can have  a much larger window of violation than a single LGI, as illustrated in the case of a photosynthetic complex in section \secref{SEC:photosynth}.  However, one could argue that measuring a combination of different LGIs will also reveal the full range of violation as this witness.
In addition, there is an extra difficulty in that testing this  witness in some cases requires the measurement of a larger number of correlation functions between all possible states in the system's Hilbert space.  Li~\etal~\cite{Cheming}  pointed out, however, that this was not strictly necessary as, since all terms on the right hand side of \eq{Qwit} are positive, one can simply truncate the summation once the right hand side is larger than the left.
Finally, Li \etal \cite{Cheming} considered the implications of an additional Markovian assumption on this equality, and showed that the resulting time-translational invariance allows one to construct a new witness which relies on state-measurements alone, and does not require the measurement of any two-time correlation functions.  However, as with the inequality of Refs.~\cite{Huelga1995,Huelga1996,Huelga1997,Waldherr2011,Zhou2012,Lambert2010}, classical non-Markovian phenomena can cause a false detection, and may be difficult to rule out.

A temporal version of Hardy's paradox has also been considered \cite{Fritz2010} that has been tested in experiment \cite{Fedrizzi2011}. Let $P(r,s|k,l)$ be the probability that Alice and Bob, measuring one after the other, obtain results $r$ and $s$, given that they chose detector settings $a_k$ and $b_l$, respectively. The (temporal) Hardy's paradox is then that the probabilities
\beq
  P(+1,+1|1,1)=0
  ;\quad
  P(-1,+1|1,2)=0
  ;\nonumber\\
  P(+1,-1|2,1)=0
  ;\quad
  P(+1,+1|2,2)>0
  \label{temporalHardy}
  ,
\eeq
as calculated under the classical assumptions \A{1-3}, are mutually inconsistent and yet, when calculated quantum-mechanically, they can indeed be simultaneously fulfilled. Both this paradox as well as the temporal CHSH of \eq{temporalCHSH} were tested with photon-polarisation qubits and results consistent with quantum-mechanics were observed \cite{Fedrizzi2011}.

\section{Conclusions\label{SEC:concs}}

The experiments discussed in this review show that we are within the era of LGI tests on {\em microscopic} systems.  The timing of this is a consequence of the  developments in quantum-computation technology over the last decade or so that have made the precise preparation and control of individual quantum systems possible.

These experiments have explored a number of interesting aspects of LGIs, such as different measurement strategies, the connection with weak values, and the effects of decoherence, etc.
However, it really comes as no surprise to find that these systems violate the LGIs.
Years of hard work in pursuit of practical quantum computation have made these systems resemble the macroscopic world as little as possible.

Despite the excellent agreement between quantum theory and experiment, if we are serious about using the LGIs to test whether a realistic (macroscopic or otherwise) description of the world is tenable, then of all the LGI tests performed to date, the only one that would cause a devout macrorealist to think twice is that of Knee~\etal~\cite{Knee2012}, since this is the only experiment to take any kind of precaution against the clumsiness loophole (Katiyar~\etal~\cite{Katiyar2013} do also consider ideal negative measurements, but their experiment is subject to other serious loopholes).
Given the fundamental requirement that the measurement operations must be perceived as being  non-invasive in order to draw any useful conclusions from a LGI violation, it is strange that only these two experiments have taken efforts to ensure this is the case. It is hard to explain why this is so, but perhaps a mistaken belief that weak measurement provide inoculation against the clumsiness loophole is partially to blame.
Of course, the measured results in all these experiments match very well the predictions of quantum theory without any nefarious detector back-action effects.  But, unless the possibility of such effects is excluded by e.g. an \inm scheme, a macrorealist can always resort to such effects to explain the results and the significance of the violations of the LGI is lost.
The analogy with the Bell inequalities is that it is no good claiming the overthrow of local-hidden-variable theories when the two parties are still at liberty to signal their results to one another.

Thus, it is clear that we are only at the outset of the journey in testing the penetration of quantum coherence into the macroscopic world with LGIs. Further progress involves not only moving up in scale to address ever-more macroscopic entities, but also in confronting the challenges posed by the clumsiness loophole.

\ack We are grateful to A.~Alberti, W.~Alt,  M.~Arndt, A.~Bednorz, W.~Belzig, \v{C}.~Brukner, C.~Budroni, T.~Calarco, J.~Dressel, E.~Gauger, M.~Goggin, G.~Knee, A.~Kofman, S.~Huelga, D.~Meschede, R.~Onofrio,
and P.~Samuelsson for their comments and suggestions.
N.~L. acknowledges the hospitality of the  Controlled Quantum Dynamics Group at Imperial College.
F.~N. is
partially supported by the 
ARO, 
RIKEN iTHES Project,
MURI Center for Dynamic Magneto-Optics, 
JSPS-RFBR Contract No. 12-02-92100, 
Grant-in-Aid for Scientiﬁc Research (S), 
MEXT Kakenhi on Quantum Cybernetics, 
and the JSPS via its FIRST Program.

\section*{References}
\bibliographystyle{prsty}

\end{document}